\documentclass[12pt]{amsart}
\usepackage{amssymb}
\usepackage{hyperref}

\newcommand\maxim[1]{\textsc{#1}}

\newtheorem{thm}{Theorem}

\newtheorem{prop}[thm]{Proposition}
\newtheorem{lemma}[thm]{Lemma}

\numberwithin{thm}{section} %% Comment out for sequentially-numbered
\numberwithin{equation}{section} %% Comment out for sequentially-numbered
\numberwithin{figure}{section} %% Comment out for sequentially-numbered


\def\CMP{Commun.\ Math.\ Phys.}

\newcommand{\bref}[1]{{\bf \ref{#1}}}

\def\ih{-i\hbar}
\newcommand{\qcommut}[2]{[#1,#2]_\star}
\newcommand{\pb}[2]{\left\{{}#1{},{}#2{}\right\}}
\newcommand{\commut}[2]{[#1,#2]}
\newcommand{\gh}[1]{{\rm gh}(#1)}
\newcommand{\p}[1]{{\rm p}(#1)}
\renewcommand{\deg}[1]{{\rm deg}(#1)}
\renewcommand{\dim}[1]{{\rm dim}(#1)}

\def\st-sig{\star_{\aA^\Sigma}}
\renewcommand{\:}{{\rm\, :\,}}
\def\bar{\overline}
\newcommand{\func}[1]{{{\mathcal C}^\infty}{(#1)}}             %
\def\tensor{\otimes}
\def\d{\partial}

\newcommand{\bl}{{\cdot}}

\newcommand{\dr}[1]{\frac{{\stackrel{\leftarrow}{\d}}}{ \d #1}}
\newcommand{\dl}[1]{\displaystyle\frac{{\d}}{\d #1}}
\newcommand{\sdl}[1]{\frac{{\d}}{\d #1}}
\newcommand{\ddl}[2]{\displaystyle\frac{{\d #1}}{\d #2}}
\newcommand{\ddd}[3]{\displaystyle\frac{{{\d}^2 #1}}{\d #2 \d #3}}

\def\half{{\frac{1}{2}}}

\def\cP{{\mathcal P}}
\def\cc{{\mathcal C}}
\def\cF{{\mathcal F}}
\def\cB{{\mathcal B}}

\def\mod{{\mathcal T}^*_\omega}

\def\manM{{\mathcal M}}
\def\manN{{\mathcal N}}

\def\aA{{ \mathfrak A}}
\def\qA{{\hat{\mathfrak A}}}
\def\E{{ \mathcal E}}

\def\P{{ \bf P}}
\def\G{{ \bf G}}

\def\V{{\bf V}}
\def\W{{\bf W}}

\def\con{{\bar\Gamma}}
\def\diff-con{{\bar\nabla}}

\def\mcon{{\boldsymbol\Gamma}}
\def\mR{\mathcal{\boldsymbol R}}
\def\mD{\mathcal{D}}
\def\mE{{\boldsymbol{\E}}}

\def\aD{{\bar D}}
\def\mnabla{\mathfrak{\nabla}}

\def\conn{{\tilde\con}}
\def\D-con{\con^0}
\def\Dcon-d{{\bar\nabla}^0}
\def\2-form{{ \bar\omega}}

\begin{document}

\thispagestyle{empty}
\vfuzz1.6pt

\baselineskip=18pt
\addtolength{\parskip}{2pt}
\raggedbottom

{\hfill{\lowercase{\tt hep-th/0101089}}\\[12pt]}
\title{Star Product for Second Class
Constraint Systems from a BRST Theory}
\author{I.A.~Batalin}
\address{I.~A.~B. \quad Lebedev Physics Institute, Russian Academy of Sciences}
\author{M.A.~Grigoriev}
\address{M.~A.~G. \quad Lebedev Physics Institute, Russian Academy of Sciences}
\author{S.L.~Lyakhovich}
\address{S.~L.~L.  \quad Department of Physics, Tomsk State University}

\begin{abstract}
  We propose an explicit construction of the
  deformation quantization of the general second-class constrained
  system, which is covariant with respect to local coordinates on
  the phase space. The approach is based on constructing the effective first-class
  constraint (gauge) system equivalent to the original second-class
  one and can also be understood as a far-going generalization of the
  Fedosov quantization.  The effective gauge system is quantized by
  the BFV--BRST procedure.  The star product for the Dirac bracket is
  explicitly constructed as the quantum multiplication of BRST
  observables.  We introduce and explicitly construct a Dirac bracket
  counterpart of the symplectic connection, called the \textit{Dirac
    connection}.  We identify a particular star product associated
  with the Dirac connection for which the constraints are in the
  center of the respective star-commutator algebra. It is shown that
  when reduced to the constraint surface, this star product
  is a Fedosov star product on the constraint surface considered
  as a symplectic manifold.
\end{abstract}
\maketitle
\thispagestyle{empty}

\vspace{0.5cm}

\begin{flushright}
\begin{minipage}{400pt}
\baselineskip=18pt
It is an honor for us to
contribute an article to the issue
commemorating the seventy-fifth birthday
of Vladimir Yakovlevich Fainberg. For many years
the authors had the priceless privililege
of being in human and scientific contact
with Vladimir Yakovlevich, and we regard 
this as a great piece of luck. Professor
V.Ya.Fainberg is permanently interested
in fundamental problems of quantum 
theory.
\end{minipage}
\end{flushright}

\vspace{0.5cm}

\section{Introduction}

In this paper we consider quantization of
second class constraint systems on general symplectic manifolds.
We work at the level of deformation quantization, i.e. our aim
is to construct the quantum multiplication (so-called star product)
in the algebra of quantum observables of the system.

The deformation approach to the quantization problem that originates
from Beresin's work~\cite{[Berezin]} was systematically formulated
in~\cite{[BFFLS]}. Existence of the deformation quantization for
arbitrary symplectic manifold was shown~\cite{[DWL]}.
An explicitly covariant procedure of constructing and
analysing star products on general symplectic manifolds,
known as Fedosov quantization, was proposed in~\cite{[Fedosov-JDG]}
(see also~\cite{[Fedosov-book]}).

In the recent paper~\cite{[GL]} it was shown that the Fedosov deformation
quantization can be understood as a BFV-BRST quantization~\cite{[BFV]}
(see also~\cite{[HT]} for a review) of the appropriately constructed
effective gauge system.  It was demonstrated that an arbitrary
symplectic manifold can be embedded as a second-class constraint
surface into the appropriately extended phase space.
By introducing additional degrees of freedom
these second class constraints can be converted
in the explicitly covariant way into the first-class
ones, giving thus the effective first-class constraint (gauge)
system that is equivalent to the original symplectic
manifold.  The BRST quantization of the system reproduces
the Fedosov star product as a quantum multiplication of
BRST observables.

An advantage of this approach is that it can be naturally generalized
to the case of constraint systems on arbitrary symplectic manifolds.
The point is that one can treat the constraints responsible for the
embedding of the phase space as a second-class surface and the
original constraints present in the model on equal
footing, which was observed relatively long
ago~\cite{{[BF89]},{[BFF]}}.  In this paper, we present a quantization
scheme for a general second-class constraint system on an
arbitrary symplectic manifold.

At the level of deformation quantization the main ingredient of the
second-class system quantization  is that of finding a star-product
that in the first order in $\hbar$ coincides with the respective
\textit{Dirac bracket}.  {}From the mathematical standpoint this
implies quantization of Poisson manifolds, with the Poisson structure
being a Dirac bracket associated to the second-class constraints.

The existence of deformation quantization for an arbitrary Poisson
manifold has been established by Kontsevich~\cite{[Kontsevich]}.  The
Kontsevich quantization formula has also been given an interesting
physical explanation in~\cite{[CF]}.  However, this formula is not
explicitly covariant\footnote{After this paper was finished we became aware of
  recent paper~\cite{[CFT]} where a globally defined version of the
  Kontsevich quantization had been proposed.} w.r.t. the phase space
coordinates, and seems too involved to apply to the case of regular
Poisson structures (the Dirac bracket for a second-class system is a
regular Poisson bracket).  In the case of regular Poisson manifolds,
one can also use an appropriate generalization of the Fedosov
quantization method~\cite{[Fedosov-book]}.  However, this scheme
requires an explicit separation of symplectic leaves of the Poisson
bracket, which in the constraint theory case implies solving the
constraints.

Unlike the system with degenerate bracket, inequivalent
observables of the second-class system are functions on the constraint
surface.  Quantization of the system implies thus not only constructing
a star product for the Dirac bracket but also an appropriate
specification of the space of quantum observables.

It turns out that within the BRST theory approach
developed in this paper one can find an explicitly covariant
quantization of a second-class system with the phase space being
an arbitrary symplectic manifold.  The correct space of observables is
described as a ghost number zero cohomology of the appropriately
constructed BRST charge.  This approach produces, in particular, a
covariant star-product for the respective Dirac bracket.

Constructing the phase space covariant quantization requires
introducing an appropriate connection on the phase space, which is
compatible with the Poisson structure to be quantized.  In the
case of an unconstraint system on a symplectic manifold
an appropriate connection, called the \textit{symplectic connection},
is a symmetric connection compatible with the symplectic form.
Together with the symplectic form on the phase space, the symplectic
connection determines a \textit{Fedosov structure}~\cite{[GRS]},
which is a basic starting point of the Fedosov quantization.

We show in this paper that by developing the BRST description of the
second-class system quantization one naturally finds a proper
Dirac counterpart of the Fedosov structure.  Namely,
the symplectic structure on the extended phase space of the effective gauge system
naturally incorporates a symmetric connection compatible with the
Dirac bracket.  This connection, called the \textit{Dirac connection},
is explicitly constructed in terms of the constraint functions and an
arbitrary symplectic connection on the phase space.

An essential feature of the Dirac connection is that
it determines a symplectic connection on the constraint
surface considered as a symplectic manifold.
Using the Dirac connection allows us to identify a star product
compatible with the constraints in the sense that the constraints are
the central functions of the respective star-commutator.
We also develop a reduction procedure which shows
that when reduced to the constraint surface this star product
can be identified as the Fedosov one, constructed by the restriction
of the Dirac connection to the surface.

In Section~\bref{sec:preliminaries}
we recall the basics of the second-class system theory and introduce
the notion of the Dirac connection. We also collect there some general
geometrical facts that we need in what follows. In
Section~\bref{sec:cl} we construct at the
classical level the effective gauge system
equivalent to the original second-class system.
Quantization of the effective system is obtained
in Section~\bref{sec:q}. The star product that determines
a quantum deformation of the Dirac bracket is also
constructed and analyzed there.
In Section~\bref{sec:reduction} we develop
the reduction of the extended phase space of the effective gauge
system to that constructed for the original constraint surface.
Finally, in Section~\bref{sec:alternative} we propose an alternative
approach to the second-class system quantization, which, in
particular, can be thought of as an alternative form of the conversion
procedure for second-class constraints.

\section{Preliminaries}\label{sec:preliminaries}
In this section we recall basic facts concerning the second 
class constraint systems on general symplectic manifolds. 
In addition to the standard facts, we propose a Dirac bracket
counterpart of the Fedosov geometry.  This includes constructing
a symmetric connection compatible with the Dirac bracket that can be
reduced to the constraint surface, thereby determining a symmetric
symplectic connection on the surface. These geometrical structures are
essential for constructing covariant star product for the Dirac
bracket.

\subsection{General phase space}
The phase space of a general second-class system is a
symplectic manifold $\manM$ with the symplectic form $\omega$
(closed and nondegenerate 2-form).  In local coordinates
$x^i\,,i=1,\ldots,2N$ on $\manM$ the coefficients of $\omega$ are
\begin{equation}
\label{aaa}
  \omega_{ij}=\omega({\frac{\d}{\d x^i},\frac{\d}{\d x^j}})\,.
\end{equation}
The Poisson bracket induced by the symplectic form reads as
\begin{equation}
  \pb{f}{g}_\manM=\ddl{f}{x^i}\omega^{ij}\ddl{g}{x^j}\,, \qquad
  \omega^{il}\omega_{lj}=\delta^{i}_{j}\,.
\end{equation}
Because we are interested in quantizing of $\manM$ we specify a
symmetric symplectic connection $\Gamma$ on $\manM$. The
compatibility condition reads as
\begin{equation}
  \nabla \omega =0 \,, \qquad \d_j\omega_{ik}-\Gamma_{ijk}+\Gamma_{kji}=0\,,
\end{equation}
where the coefficients $\Gamma_{jik}$ are introduced as
\begin{equation}
  \Gamma_{jik}=\omega_{jl} \Gamma^l_{ik}\,, \qquad
  \Gamma^j_{ik}\dl{x^j}=\nabla_{i}\dl{x^k}\,.
\end{equation}
Given a symplectic form and a compatible symmetric connection on
$\manM$ one says that $\manM$ is a \textit{Fedosov
  manifold}~\cite{[GRS]}.  It is well known that a symmetric symplectic
connection exists on every symplectic manifold, each
symplectic manifold is therefore a Fedosov one. However,
unlike the Levi-Civita connection, a symmetric symplectic
connection is not unique; the arbitrariness is that of adding
a completely symmetric tensor field to the coefficients
$\Gamma_{jik}$. Thus the Fedosov structure is not completely
determined by the symplectic one.

\subsection{Second-class constraint system}
Let now $\manM$ be a phase space of a
second-class constraint system specified by the following set of
the second-class constraints:
\begin{equation}
\theta_\alpha=0\,,\qquad \alpha=1,\ldots\,,2M\,.
\label{eq:constraints}
\end{equation}
We assume $\theta_\alpha$ to be globally defined functions on $\manM$.
Let $\Sigma \subset \manM$ denotes the respective constraint
surface.  The Dirac matrix reads as
\begin{equation}
  \label{eq:dirac-matrix}
\Delta_{\alpha \beta}=\pb{\theta_\alpha}{\theta_\beta}
\end{equation}
and is assumed to be invertible. Its inverse is denoted by
$$
\Delta^{\alpha \beta}\,,\qquad
 \Delta_{\alpha \gamma}\Delta^{\gamma \beta}=\delta^\beta_\alpha\,.
$$
Invertibility of the Dirac matrix implies, at the same time,
that $\Sigma$ is a smooth submanifold in $\manM$.

\textit{Inequivalent observables} of the second-class system
are functions on the constraint surface $\Sigma$.  The algebra of
inequivalent observables of the second-class system on $\manM$ is a
Poisson algebra of functions on the constraint surface, with the
Poisson structure corresponding to the symplectic form
$\omega^\Sigma\equiv\omega{\big|}_{\Sigma}$
(the 2-form $\omega{\big|}_{\Sigma}$ denotes the restriction of
the 2-form $\omega$ to $\Sigma\subset\manM$;
$\omega{\big|}_{\Sigma}$ is nondegenerate since $\Delta_{\alpha\beta}$
is).  At the classical level, two constraint systems are said
equivalent iff their algebras of inequivalent observables are
isomorphic as Poisson algebras.

\subsection{Dirac bracket}

The goal of the Dirac bracket approach to the second-class system
is to represent the algebra of inequivalent observables as a quotient
of the algebra of functions on $\manM$ modulo the ideal generated by
the constraints. The point is that $\manM$ can be equipped with a
Poisson bracket called the \textit{Dirac bracket} which is well defined on
this quotient.  Given second-class constraints $\theta_\alpha$ the
respective Dirac bracket has the following form:
\begin{equation}
  \label{eq:Dirac}
\pb{f}{g}^D_\manM=\pb{f}{g}_\manM-
\pb{f}{\theta_\alpha}_\manM \Delta^{\alpha \beta} \pb{\theta_\beta}{g}_\manM\,.
\end{equation}
We recall the basic properties of a Dirac bracket:
\begin{equation}
\label{eq:DB-properties}
  \begin{split}
\pb{f}{g}^D_\manM+\pb{g}{f}^D_\manM&=0\\
\pb{f}{gh}^D_\manM-\pb{f}{g}^D_\manM h- g \pb{f}{h}^D_\manM& =0\\
\pb{f}{\pb{g}{h}^D_\manM}^D_\manM+{\rm cycle}(f,g,h)&=0\\
\pb{f}{\theta_\alpha}^D_\manM&=0\,.
\end{split}
\end{equation}
The first three lines say that $\pb{}{}^D_\manM$ is a Poisson
bracket.  In what follows we use the notation
$D^{ij}$ for the components of the Poisson bivector of the Dirac bracket:
\begin{equation}
D^{ij}=\pb{x^i}{x^j}^D_\manM\,.
\end{equation}
The last line in \eqref{eq:DB-properties} implies that $\theta_\alpha$
are characteristic  (Casimir) functions of the bracket
$\pb{}{}^D_\manM$. This, in turn, implies that
$\pb{}{}^D_\manM$ is well-defined on the quotient
algebra of $\func{\manM}$ modulo functions of the form
$F^\alpha\theta_{\alpha}$ (i.e. modulo the ideal
of functions vanishing on $\Sigma$). This quotient can be naturally
identified with the algebra $\func\Sigma$ of functions on $\Sigma$.
Thus $\func\Sigma$ is a Poisson algebra and $\Sigma$
is a Poisson manifold. We refer to the Poisson
bracket in $\func\Sigma$ as to the restriction of a Dirac bracket
to $\Sigma$.

The constraint surface $\Sigma$ is in fact a symplectic manifold,
with the symplectic form $\omega^\Sigma\equiv \omega\big|_{\Sigma}$.
The Poisson bracket on $\Sigma$ corresponding to the symplectic form
$\omega^\Sigma$ coincides with the restriction
$\pb{}{}^D_\manM{\bigr|}_\Sigma$ of a Dirac bracket to $\Sigma$.
Thus description of a second-class system in terms
of a Dirac bracket is equivalent to that based on the
reduction to the constraint surface.

From the geometrical viewpoint, $\theta_\alpha$ give a
maximal set of independent characteristic functions of the Dirac
bracket.  This implies that $\manM$ is a regular Poisson manifold.
Each surface of the constant values of the functions
$\theta_\alpha$ is therefore a symplectic leaf of the Dirac bracket
and is a symplectic manifold. In particular, constraint surface
$\Sigma$ is a symplectic leaf.  As we have seen the symplectic form
$\omega|_{\Sigma}$ on $\Sigma$ (on each symplectic leaf) is the
restriction of the symplectic form $\omega$ on $\manM$ to
$\Sigma \subset \manM$ (respectively the symplectic leaf).

\subsection{Dirac connection}\label{subsec:D-connection}
At the level of quantum description we also need the
\textit{Dirac connection} that is a symmetric
connection compatible with the Dirac bracket.  Given a symmetric
symplectic connection $\Gamma$ on $\manM$ there exists a Dirac
connection $\con^0$ whose coefficients are
\begin{equation}
\label{eq:Dirac-con}
{(\con^0)}^k_{ij}=\omega^{kl}
(\Gamma_{lij}+ \d_l \theta_\beta \Delta^{\beta \alpha} \nabla_i \nabla_j \theta_\alpha
)\,,\qquad \qquad \nabla_i \nabla_j \theta_\alpha= \d_i\d_j
\theta_\alpha-\Gamma^m_{ij}\d_m \theta_\alpha\,,
\end{equation}
where $\nabla_i$ is the covariant derivative w.r.t. $\Gamma$.  It is
a matter of direct observation that $\con^0$ preserves the Dirac
bivector: $\Dcon-d_i D^{jk}$ where the notation $\Dcon-d$ is
introduced for the covariant differentiation determined by
the connection $\con^0$. Moreover, one can check that
$\Dcon-d_i \d_j \theta_\alpha=0$.  The curvature
of the Dirac connection is given explicitly by
\begin{multline}
    ({\bar R}^0)^m_{ij;\,l}~=~D^{mk}{\bar R}^0_{ij;\,kl}~=\\
=~D^{mk}\left(R_{ij\,;kl}+
(\nabla_{i}\nabla_{k}\theta_\alpha)
\Delta^{\alpha\beta}
(\nabla_{j}\nabla_{l}\theta_\beta)-
(\nabla_{j}\nabla_{k}\theta_\alpha)
\Delta^{\alpha\beta}
(\nabla_{i}\nabla_{l}\theta_\beta)\right)\,,
\end{multline}
where $R_{ij\,;kl}=\omega_{km}R^m_{ij;\,l}$ is the curvature of
$\Gamma$.

An important point is that the Dirac connection $\con^0$ on $\manM$
determines a connection on the constraint surface $\Sigma$.
This implies that the parallel transport along $\Sigma$ carries
vectors tangent to $\Sigma$ to tangent ones and thus determines a
connection on
$\Sigma$. 
To show that $\con^0$ does restrict to $\Sigma$ let $Y,Z$ be vector
fields on $\manM$ tangent to $\Sigma$.  This implies that
there exist functions $Y^\alpha_\beta,Z^\alpha_\beta$ such that
\begin{equation}
  Y\theta_\alpha=Y^\beta_\alpha\theta_\beta\,, \qquad
  Z\theta_\alpha=Z^\beta_\alpha\theta_\beta\,.
\end{equation}
Let us show that $\Dcon-d_Y Z$ is also
tangent to $\Sigma$. Indeed,
\begin{multline}
    (\Dcon-d_Y Z)\theta_\alpha=
    (Y^i\d_iZ^j)\d_j \theta_\alpha+Y^i(\con^0)^j_{ik}Z^k\d_j\theta_\alpha=\\
=    YZ\theta_\alpha-Y^iZ^j\d_i\d_j \theta_\alpha
+Y^iZ^k \Gamma^j_{ik} \d_j \theta_\alpha
+Y^iZ^k  \omega^{jn} \d_n \theta_\gamma \Delta^{\gamma\beta}
\nabla_i\nabla_k\theta_\beta \d_j \theta_\alpha=\\
=YZ\theta_\alpha=(YZ^\beta_\alpha)\theta_\beta+Z^\beta_\alpha
Y^\gamma_\beta \theta_\gamma\,.
\end{multline}
The last expression obviously vanishes on $\Sigma$. The vector field
$\Dcon-d_Y Z$ is therefore tangent to $\Sigma$.  This implies that
connection $\con^0$ on $\manM$ determines a connection $(\con^0)^\Sigma$ on
$\Sigma$.  Since $\con^0$ preserves Dirac bivector $D^{ij}\sdl{x^i}
\wedge \sdl{x^j}$, reduced connection $(\con^0)^\Sigma$ preserves its
restriction to $\Sigma$.  Because the Poisson bracket corresponding to the
symplectic form $\omega^\Sigma=\omega\bigr|_{\Sigma}$ is the restriction
of the Dirac bracket to $\Sigma$ , $(\con^0)^\Sigma$ is a symmetric
symplectic connection on $\Sigma$.  Thus any second-class constraint
surface in the Fedosov manifold is also a Fedosov manifold, with the
symplectic structure and a compatible symmetric connection being
the restrictions to the constraint surface of the symplectic structure and
the Dirac connection on the phase space respectively.

Let us also consider a coordinate description for the reduction
of a Dirac connection to the constraint surface.  To this end, we
note that expression ~\eqref{eq:Dirac-con} for the Dirac connection
can be equivalently rewritten as:
\begin{equation}
{(\con^0)}^k_{ij}=D^{kl}\Gamma_{lij}+
\omega^{kl}\d_l \theta_\beta \Delta^{\beta\alpha} \d_i \d_j \theta_\alpha \,.
\end{equation}

Further, let us pick functions
${\bar x}^a\,,~~a=1,\ldots, 2N-2M$
such that ${\bar x}^a$ and
${\bar x}^\alpha=\theta_\alpha\,,~~\alpha=1,\ldots, 2M$ form a local
coordinate system on $\manM$.  This coordinate system
is adopted to the embedded submanifold $\Sigma \subset \manM$
in the sense that locally $\Sigma$ is singled out by equations
${\bar x}^\alpha=0$.  It follows that the functions
${\bar x}^a|_{\Sigma}\,,~~a=1,\ldots, 2N-2M$ form
a local coordinate system on $\Sigma$.  In the coordinate system
${\bar x}^a\,,{\bar x}^\alpha$ on $\manM$ the expression
$\d_i\d_j \theta_\alpha$ obviously vanishes and the Dirac connection
takes the form
\begin{equation}
{(\con^0)}^k_{ij}=D^{kl}\Gamma_{lij}\,,
\end{equation}
where we use $i$ (or $j,k,l)$ as a collective notations for either $a$
or $\alpha$.  Because
$D^{i\alpha}=\pb{{\bar x}^i}{{\bar x}^\alpha}^D_{\manM}=0$,
the respective components of the Dirac connection
vanishes:
\begin{equation}
\label{eq:DGamma-vanish}
  (\con^0)^\alpha_{ij}=0\,.
\end{equation}
Then, for a vector field tangent to $\Sigma$ one has
\begin{equation}
\nabla_{\frac{\d}{\d {\bar x}^a}}\dl{{\bar x}^b}=
(\con^0)^l_{ab}\dl{{\bar x}^l}=
(\con^0)^c_{ab}\dl{{\bar x}^c}\,,
\end{equation}
in view of~\eqref{eq:DGamma-vanish}.  When restricted to $\Sigma$
the equation implies that the functions $(\con^0)^a_{bc}{\big|}_{\Sigma}$
are the coefficients of a symmetric symplectic connection
$(\con^0)^\Sigma$ on $\Sigma$ w.r.t. the coordinate system
${\bar x}^a{\bigr|}_\Sigma$ on $\Sigma$.

\subsection{Modified Poisson bracket on the cotangent bundle over
a symplectic manifold}\label{subsec:mod}
The phase space $\manM$ of the system under consideration is a
general symplectic manifold.  Even without constraints the
quantization problem for $\manM$ can not be solved directly and
requires using specific quantization methods.   Since we are interested
in quantizing a constraint system the most suitable method is that
based on the representation of $\manM$ itself as a second-class
constraint system; the method was proposed in \cite{[BF89]} and
then generalized to the case of arbitrary symplectic manifolds in
\cite{[GL]}.  It was also shown to reproduce the Fedosov quantization
in terms of the BRST formulation of the constraint system theory.

In order to represent $\manM$ as a second-class constraint surface let us
consider first an appropriate generalization of the canonical
Poisson bracket on a cotangent bundle.  Let $T^*\manM$ be a
cotangent bundle over $\manM$ and $x^i,p_j$ be standard
coordinates on the base and the fibers respectively.  Let also
$\pi\: T^*\manM \to \manM$ be a bundle projection which sends a
1-form at point $m$ to $m$ (points of
$T^*\manM$ are pairs $(m,a)\: m \in \manM,\, a \in T^*_m \manM$)
and $\pi^* \: \Lambda(\manM) \to \Lambda(T^*\manM)$
is the pullback map associated with $\pi$.
Here we use notation
$\Lambda(\manM)=\oplus_{k=0}^{\dim{\manM}}\Lambda^k(\manM)$
for the space of differential forms on $\manM$.

Cotangent bundle $T^*\manM$ is equipped with the canonical symplectic
structure:
$$
\omega^{T^*\manM}=2\,dp_i \wedge dx^i
$$
The corresponding Poisson bracket also has standard
form:
$$
  \pb{x^i}{p_j}_{T^*\manM}=\delta^i_j\,.
$$
Let $\omega=\omega_{ij} dx^i \wedge dx^j$ be a closed 2-form on
$\manM$.  Let us introduce a modified symplectic structure on
$T^*\manM$ by:
\begin{equation}
  \label{eq:mod-symplectic}
\omega^{mod}=
\omega^{T^*\manM}+\pi^* \omega=2\,dp_i \wedge dx^j + \omega_{ij} dx^i
\wedge dx^j\,.
\end{equation}
This 2-form is obviously nondegenerate and closed.
Hereafter $\mod\manM$ denotes cotangent bundle $T^*\manM$
equipped with the modified symplectic structure.
Symplectic structure $\omega^{mod}$ determines a Poisson bracket
on $\mod\manM$, with the basic Poisson bracket relations given by:
\begin{equation}
  \label{eq:mod-PB}
\pb{x^i}{x^j}_{\mod\manM}=0\,, \qquad
\pb{x^i}{p_j}_{\mod\manM}=\delta^i_j\,, \qquad
\pb{p_i}{p_j}_{\mod\manM}=\omega_{ij}(x)\,.
\end{equation}
The Jacobi identity for this bracket holds provided 2-form
$\omega$ is closed.

One can easily check that (at least locally) one can bring the bracket
to the standard (canonical) form by means of the following transformation
$p_i \to p_i-\rho_i(x)$, with $\rho_i(x)$ being a symplectic
potential for the symplectic structure $\omega$:
\begin{equation}
  \omega=2\, d\rho\,, \qquad \omega_{ij}=\d_i \rho_j -\d_j\rho_i\,.
\end{equation}

Until this point 2-form $\omega$ could be either nondegenerate
or not. Let now $\omega_{ij}$ be a symplectic structure of
$\manM$. The Poisson brackets of the coordinate functions
$p_i$ form then an invertible matrix. Constraints
\begin{equation}
\theta_i=0\,,\qquad \theta_i\equiv -p_i
\end{equation}
(we choose minus sign for convenience) are then the second-class
ones, with the respective constraint surface being $\manM$
(considered as a zero section of $\mod\manM$).  In
general, these constraints are not the globally defined functions on
$\mod\manM$.  This, in particular, implies that the Dirac bracket
associated with constraints $\theta_i$ is not a globally defined
Poisson bracket on $\mod\manM$.  However, this Dirac bracket is well
defined on the constraint surface $\manM \subset \mod\manM$.  When
restricted to constraint surface $\manM$ this Dirac bracket coincides with
the Poisson bracket $\pb{}{}_\manM$ on $\manM$. In this way one can
represent an arbitrary Hamiltonian system on a symplectic manifold
as the second class constraint system on the modified
cotangent bundle over the manifold.

\subsection{Symplectic structure}\label{subsec:symplectic}
In what follows we also need a specific Poisson bracket
on the appropriately extended cotangent bundle over a symplectic
manifold.   Given an arbitrary symplectic vector
bundle $\W(\manN) \to \manN$ over a manifold $\manN$ let $e_A$ be
a local frame (locally defined basic sections of $\W(\manM)$).  Let
also $\mD$ be the symplectic form on the fibres of $\W(\manM)$.  The
components of $\mD$ w.r.t. $e_A$ are determined by
$\mD_{AB}=\mD(e_A,e_B)$.

It is well known (see e.g.~\cite{[Fedosov-book]}) that any symplectic
vector bundle admits a symplectic connection.  Let $\mcon$ and
$\mnabla$ denotes a symplectic connection and the corresponding covariant
differential in $\W(\manN)$. The compatibility condition reads as
\begin{equation}
\label{eq:compatible}
  \mnabla \mD=0\,,  \qquad \d_i \mD_{AB} - \mcon^C_{iA}\mD_{CB}-\mcon^C_{iB}\mD_{AC}=0\,,
\end{equation}
where the coefficients $\mcon^C_{iA}$ of $\mcon$ are determined
as:
\begin{equation}
  \mnabla e_A=dx^i \mcon^C_{iA}e_C\,.
\end{equation}
It is useful to introduce the following connection 1-form:
\begin{equation}
\mcon_{AB}=dx^i \mcon_{AiB}\,, \qquad  \mcon_{AiB}=\mD_{AC}\mcon^C_{iB}\,.
\end{equation}
Then compatibility condition \eqref{eq:compatible} rewrites as
\begin{equation}
\label{eq:compatible-l}
d\mD_{AB}=\mcon_{AB}-\mcon_{BA}\,, \qquad  \d_i \mD_{AB} - \mcon_{AiB}+\mcon_{BiA}=0\,.
\end{equation}
As a consequence of the condition one arrives at the following
property of the connection 1-form $\mcon_{AB}$:
\begin{equation}
\label{eq:dC-symm}
  d\mcon_{AB}=d\mcon_{BA}\,.
\end{equation}

Consider the following direct sum of vector bundles:
\begin{equation}
  \mE=\W(\manN) \oplus \mod\manN\,,
\end{equation}
where for generality we assume that $\manN$ is equipped with a closed
2-form $\omega$ and $\mod\manN$ is a modified cotangent bundle
over $\manN$.  Let $x^i,p_j$ and $Y^A$ are standard local coordinates
on $\mE$ ($x^i$ are local coordinates on $\manN$, $p_j$ are standard
coordinates on the fibres of $\mod\manN$, and $Y^A$ are coordinates on
the fibres of $\W(\manN)$ corresponding to the local frame $e_A$).

We claim that considered as a manifold $\mE$ is equipped with the following
symplectic structure
\begin{equation}
  \begin{split}
  \label{eq:symplectic}
  \omega^{\,_{\mE}}~~=~~&\pi^*\omega+2dp_i\wedge dx^i+\\
+~&\mD_{AB}dY^A\wedge
  dY^B+ Y^AY^B d\mcon_{AB} -2Y^A\mcon_{AB}\wedge dY^B\,,
\end{split}
\end{equation}
where $\pi^* \omega$ is the 2-form $\omega$
on $\manN$ pulled back by bundle projection $\pi\:\mE \to \manN$.  One
can directly check that 2-form \eqref{eq:symplectic} is well defined.
That it is closed follows from $d \omega=0$,
condition~\eqref{eq:compatible-l}, and Eq.~\eqref{eq:dC-symm}.

The Poisson bracket on $\mE$ corresponding to the symplectic
form~\eqref{eq:symplectic} is determined by the following basic relations:
\begin{equation}
  \begin{aligned}
    \pb{x^i}{p_j}_{\mE}&=\delta^i_j\,,\\
    \pb{Y^A}{Y^B}_{\mE}&=\mD^{AB}(x)\,,\\
    \pb{Y^A}{p_i}_{\mE}&= - \mcon^A_{iB}(x)Y^B\,,\\
    \pb{p_i}{p_j}_{\mE}&=\omega_{ij}(x)+\half \mR_{ij;AB}(x)Y^A\,Y^B\,,
  \end{aligned}
  \label{eq:extended-PB}
\end{equation}
with all the others vanishing: $\pb{x^i}{Y^A}_{\mE}=\pb{x^i}{x^j}_{\mE}=0$. Here,
$\mR_{ij;\,AB}$ denotes the curvature of $\mcon$:
\begin{multline}
  \mR_{ij;\,AB}=\mD_{AC}\mR^C_{ij\,B}~=\\
=~\mD_{AC}\left(\d_i \mcon^C_{jB}-\d_j \mcon^C_{iB}+\mcon^C_{iD}\mcon^D_{jB}
-\mcon^C_{jD}\mcon^D_{iB}\right)~=\\
  =~\d_i \mcon_{AjB}-\d_j
  \mcon_{AiB}+\mcon_{CiA}\mD^{CD}\mcon_{DjB}-\mcon_{CjA}\mD^{CD}\mcon_{DiB}\,.
\end{multline}
The last equality follows from nondegeneracy of $\mD_{AB}$
and compatibility condition~\eqref{eq:compatible-l}.

\section{Conversion -- classical description}\label{sec:cl}
In this section we construct the BFV-BRST description of the
second-class constraint system on $\manM$ in terms
of an equivalent effective first-class constraint (gauge) system.
This includes explicit construction of the gauge system,
its BRST charge, and observables.

A general procedure for conversion of the first class constraint systems into the
second class ones was proposed in~\cite{[BF],[BT],[BFF]} (see
also~\cite{[FL]}). In these papers the phase space of the second class 
system is extended by linear symplectic space (as multiplication by a
factor) and the conversion variables are introduced as coordinates on
this linear factor.  In this paper we use slightly modified
conversion scheme, where conversion variables are introduced as
coordinates on the fibres of the appropriate vector bundle, associated
to the constraints. This scheme allows one to consider all the
constraints (i.e. constraints determining the embedding of the
original symplectic manifold $\manM$ into the modified cotangent
bundle and the original constraints on $\manM$) on equal footing
and keep the explicit covariance at all stages of the conversion.

\subsection{Unification of constraints $\theta_\alpha$ and $\theta_i$}
Given a second-class constraint system on $\manM$ we first embed
$\manM$ as a zero section in $\mod\manM$.  According to
Section~\bref{subsec:mod} the constraints $\theta_i\equiv -p_i=0$
represent $\manM$ as a second-class constraint
surface in $\mod\manM$ and determine on $\mod\manM$ the
constraint system which is equivalent to the original symplectic
manifold $\manM$ (i.e. an unconstraint system on $\manM$).

In order to describe the constraint system on $\manM$ specified by
the second-class constraints $\theta_\alpha$ one considers
a constraint system on $\mod\manM$, with the constraints being
$\theta_i$ and $\theta_\alpha$.  We treat these constraints on
equal footing and introduce a unified notation:
\begin{equation}
\Theta_A=(\theta_i\,,\theta_\alpha)\,,
 \qquad A=1,\ldots ,2N+2M\,.
\end{equation}
Their Dirac matrix is
\begin{equation}
  \label{eq:Dirac-Theta}
D_{AB}=\pb{\Theta_A}{\Theta_B}_{\mod\manM}\,.
\end{equation}
Explicitly, $D_{AB}$ is given by
\begin{equation}
D=
\left(\begin{array}{cc}
D_{ij} & D_{i\beta}\\[7pt]
D_{\alpha j}& D_{\alpha\beta}
 \end{array}\right)=
\left(\begin{array}{cc}
\omega_{ij} & \d_i \theta_\beta\\[7pt]
- \d_j\theta_\alpha & 0
 \end{array}\right)
\end{equation}

\medskip

\noindent
One can check that
$det(D)=det(\omega_{ij})det(\Delta_{\alpha \beta})$.
The matrix $D^{AB}\equiv(D^{-1})^{AB}$ that is inverse to $D_{AB}$ reads as
\vspace{0.3cm}
\begin{equation}
\label{eq:DAB}
D^{\,-1}= 
\left(\begin{array}{cc}
D^{ij} & D^{i\beta}\\[7pt]
D^{\alpha j}& D^{\alpha\beta}
 \end{array}\right)=
\left(
\begin{array}{cc}
\omega^{ij}+\omega^{ik}\omega^{jl} \d_k \theta_\alpha \d_l
\theta_\beta \Delta^{\alpha\beta} &\quad  - \omega^{il}\d_l \theta_\gamma
\Delta^{\gamma \beta }\\[11pt]
\omega^{jk} \d_k \theta_\gamma \Delta^{\gamma \alpha }& \quad
\Delta^{\alpha \beta}
\end{array}\right)
\end{equation}

\vspace{0.3cm}

\noindent
Note that the left upper block of the matrix is nothing but the
Poisson bivector of the Dirac bracket on $\manM$ associated with
the second-class constraints $\theta_\alpha$.

Among the constraints $\Theta_A$ there are constraints
$\theta_i$ which are not the globally defined functions on $\mod\manM$;
they transform as the components of a $1$-form on $\manM$.
Consequently the Dirac bracket associated with $\Theta_A$
is not a globally defined Poisson bracket on $\mod\manM$.
However, it is well defined for $p_i$-independent functions.
\begin{prop}
\label{prop:3.1}
Let $f$ and $g$ be arbitrary functions on $\manM$.
Let $\pi^*$ be the pullback associated with the bundle projection
$\pi \: \mod\manM\to  \manM$. Then
\begin{equation}
  \label{eq:coincidence}
\pb{\pi^* f}{\pi^* g}_{\mod\manM}^{D}=
\pi^*(\pb{f}{g}_{\manM}^D)\,,
\end{equation}
where the Dirac bracket in the L.H.S. is taken w.r.t. the
constraints $\Theta_A$ and in the R.H.S. w.r.t. $\theta_\alpha$
only. In particular, these Dirac brackets are identical on $\Sigma \subset 
\manM$.
\end{prop}
Because each physical observable of the constraint system on
$\mod\manM$ can be taken as a $p_i$-independent function, Proposition~\bref{prop:3.1}
implies that the original constraint system on $\manM$ (determined by
the constraints $\theta_\alpha$) is equivalent with the constraint
system on $\mod\manM$ (determined by the constraints $\Theta_A$).
A direct way to check the equivalence of these constraint systems is to
observe that the constraints $\Theta_A$ on $\mod\manM$  and $\theta_\alpha$
on $\manM$ determine the same constraint surface $\Sigma$ and
the respective Dirac brackets $\pb{}{}_{\mod\manM}^{D}$ and
$\pb{}{}_{\manM}^D$ coincide on $\Sigma$.

\subsection{Symplectic connection}\label{subsec:connection}
In order to convert the second-class constraints $\Theta_A$ into
the first-class ones we introduce the conversion variables $Y^A$
associated to $\Theta_A$.  We treat
the variables $Y^A$ as coordinates on the fibres of the
vector bundle $\W(\manM)$ associated to the constraints
$\Theta_A$.  This means that variables $Y^A$ have the transformation
properties dual to those of $\Theta_A$. Thus the phase space of the
extended system is given by: $\E_0=\mod\manM\oplus\W(\manM)$.

Because constraints $\Theta_A$ are split into $\theta_i$ and
$\theta_\alpha$, the vector bundle
$\W(\manM)$ is a direct sum
\begin{equation}
\W(\manM)=T\manM\oplus \V(\manM)=T\manM \times V\,,
\end{equation}
where $\V(\manM)$ is the vector bundle associated with the constraints
$\theta_\alpha$.  Since the constraints
$\theta_\alpha\,,~~\alpha=1,\ldots,2M\,$
are globally defined functions on $\manM$, the vector bundle
$\V(\manM)$ is a direct product of $\manM$ and $2M$-dimensional
vector space $V$.

The Dirac matrix $D_{AB}$ obviously determines a symplectic form on
each fiber of $\W(\manM)$, making $\W(\manM)$ into the symplectic
vector bundle.  To convert the second-class constraints $\Theta_A$
into the first-class ones one has to extend the Poisson bracket on
$\mod\manM$ to the phase space $\E_0=\mod\manM\oplus\W(\manM)$ of
the extended system.  This in turn requires to introduce a symplectic
connection in $\W(\manM)$.

Each symplectic vector bundle admits a symplectic connection. However,
it is instructive to find the explicit form of the symplectic
connection in $\W(\manM)=T\manM\oplus \V(\manM)$ compatible with
the symplectic form $D$.  Moreover, in what follows we need
the specific symplectic connection $\con^0$ in $\W(\manM)$
which is constructed below.

Introducing coefficients of $\con$ with lowered
indices as
\begin{equation}
  \con_{AiB}=D_{AC}\con^C_{iB}\,,
\end{equation}
we write equation \eqref{eq:compatible-l} in components:
\begin{equation}
\begin{split}
\dl{x^i} \omega_{jk}+\con_{kij}-\con_{jik}&=0\,,\\
\ddd{}{x^i}{x^j}{\theta_\alpha}
+\con_{\alpha ij}-\con_{ji\alpha}&=0\,,\\
\con_{\alpha i \beta}-\con_{\beta i \alpha}&=0\,,
\end{split}
  \label{eq:con-components}
\end{equation}
The first equation is that for a symplectic connection
on $\manM$. Thus it is natural to chose a particular solution
to the first equation as
\begin{equation}
\con^0_{kij}=\omega_{kl}\Gamma^l_{ij}\,,
  \label{eq:first}
\end{equation}
where $\Gamma^l_{ij}$ are coefficients of the fixed symmetric
symplectic connection on $\manM$.  Further, under the change of
coordinates on $\manM$ the coefficients
$\con_{i j \alpha}$ transform as the the components of a tensor field.
Thus the condition $\con_{ij \alpha}=0$ is the invariant one and one
can choose a particular solution of the second equation as:
\begin{equation}
\con^0_{ij \alpha}=0\,, \qquad
\con^0_{\alpha i j}=-\d_i\d_j \theta_\alpha\,.
\end{equation}
Finally, we choose $\con^0_{\alpha i \beta}=0$.  Thus
we obtain the particular solution $\con^0_{AiB}$ for
equations~\eqref{eq:con-components}. A general solution
is obviously given by
\begin{equation}
  \con_{AiB}=\con^0_{AiB}+T_{AiB}\,, \qquad T_{AiB}-T_{BiA}=0
  \label{eq:con-general}
\end{equation}
where $T_{AiB}$ is an arbitrary 1-form on $\manM$
with values in the symmetric tensor square of the bundle
$\W(\manM)$.

An explicit expression for the non-vanishing coefficients
of $\con^0$ read as:
\begin{equation}
\label{eq:con0-explicit}
\begin{split}
(\con^0)^i_{jk}&=
D^{iA}{\con^0_{Ajk}}=
D^{il}{\con^0_{ljk}}+D^{i\alpha} {\con^0_{\alpha jk}}=
\omega^{il}(\Gamma_{ljk}+\d_l\theta_\beta \Delta^{\beta \alpha}
\nabla_j \nabla_k \theta_\alpha)\,,\\
(\con^0)^\alpha_{jk}&=D^{\alpha A} \con^0_{A jk}=
D^{\alpha l} \con^0_{ljk}
+D^{\alpha \beta} \con^0_{\beta jk}=
-\Delta^{\alpha \beta}\nabla_j\nabla_k \theta_\beta
\end{split}
\end{equation}
One can see that the coefficients $(\con^0)^i_{jk}$ coincide with those of
the Dirac connection on $\manM$ given by~\eqref{eq:Dirac-con}. Indeed,
$D^{ij}$ is nothing but the Poisson bivector of the Dirac bracket on
$\manM$. In its turn, compatibility condition~\eqref{eq:compatible}
implies
\begin{equation}
\Dcon-d_i D^{jk}=\d_i D^{jk}+(\con^0)^j_{im}D^{mk}+(\con^0)^k_{im}D^{jm}=0\,,
\end{equation}
since the coefficients $(\con^0)^j_{i\alpha}$ vanish. This allows one
to consider the Dirac connection on $\manM$ as the restriction
of $\con^0$ in $\W(\manM)$ to $T\manM\subset \W(\manM)$.

Let us also write in components the curvature
of $\con^0$. The curvature tensor with lowered indices
is determined by:
\begin{equation}
  \begin{split}
    {\bar R}^0_{ij;AB}&~=~D_{AC}(R^0)^C_{ij\,B}~=\\
=~&\d_i \con^0_{AjB}-\d_j \con^0_{AiB}+\con^0_{CiA}D^{CD}\con^0_{DjB}
-\con^0_{CjA}D^{CD}\con^0_{DiB}\,.
  \end{split}
\end{equation}
The only non-vanishing components of $R^0_{ij;AB}$ are
given by
\begin{multline}
    {\bar R}^0_{ij;kl}~~=~~\d_i \con^0_{kjl}-
\d_j \con^0_{kil}+
\con^0_{Aik}D^{AB}\con^0_{Bjl}-
\con^0_{Ajk}D^{AB}\con^0_{Bil}~~=\\
=~~R_{ij;\,kl}+
(\nabla_{i}\nabla_{k}\theta_\alpha)
\Delta^{\alpha\beta}
(\nabla_{j}\nabla_{l}\theta_\beta)-
(\nabla_{j}\nabla_{k}\theta_\alpha)
\Delta^{\alpha\beta}
(\nabla_{i}\nabla_{l}\theta_\beta)\,,
\end{multline}
where $R_{ij;\,kl}$ is the Riemannian curvature of the symplectic
connection $\Gamma$ on $\manM$ from~\eqref{eq:first}.

In what follows it goes without saying that the vector bundle
associated with constraints $\Theta_A$ is a direct sum
$T\manM \oplus \V(\manM)=T\manM \times V$ and is equipped with
the symplectic connection $\con$.  We also reserve
notations $\con^0$ and ${\bar R}^0$ for the specific symplectic
connection given by \eqref{eq:con0-explicit} and its curvature.

\subsection{Extended phase space}\label{subsec:extended-phase-space}

Recall (see subsection~\bref{subsec:symplectic}) that given
a symplectic vector bundle $\W(\manN)$ endowed with a symplectic
connection $\mcon$ and a symplectic form $\mD$ one can equip
$\W(\manN)\oplus \mod\manN$ with the symplectic structure.
Specifying this general construction to the vector bundle
$\W(\manM)=T\manM \times V$ equipped with the symplectic form $D$ and
the connection $\con$ one arrives at the following symplectic
structure on the phase space $\E_0=\mod\manM\oplus\W(\manM)$:
\begin{equation}
\label{eq:symplectic-E0}
\begin{split}
  \omega^{{\,\E_0}}~~=~~\pi^*\omega~+~&  2dp_i \wedge dx^i ~~+\\
 & +~~ D_{AB}dY^A\wedge dY^B~+~Y^A Y^B d \con_{AB}~-~2\con_{AB}
  \wedge dY^AY^B\,.
\end{split}
\end{equation}

Our aim is to construct an effective first-class constraint theory
on $\E_0$ by converting the second-class constraints $\Theta_A$
into the first-class ones.  At the classical level, the conversion is
achieved by continuation of the constraints $\Theta_A$ into the new
constraints $T_A$ defined on $\E_0$ such that~\cite{[BF87],[BT]}:
\begin{equation}
  \label{eq:determin}
\pb{T_A}{T_B}_{\E_0}=0\,, \qquad T_A|_{Y^A=0}=\Theta_A\,.
\end{equation}
The constraints $T_A$ are understood as formal power series in $Y^A$
\begin{equation}
  T_A=\sum_{s=0}^{\infty}T^s_{A}\,,\qquad T^0_A=\Theta_A\,,\quad
  T^s_A=T^s_{A\, B_1,\ldots , B_s}(x)Y^{B_1},\ldots , Y^{B_s}\,,
  \label{eq:T-exp}
\end{equation}
where the coefficients $T^s_{A B_1,\ldots , B_s}$
are assumed to be $p_i$-independent functions.

In spite of the fact that the constraints $T_A$ are Abelian
by construction, it is useful to proceed within the BFV-BRST approach.
Accordingly, we introduce the  ghost variables $\cc^A$ and
$\cP_A\,, A=1\,,\ldots\,,2M+2N$ associated to the constraints $T_A$.
Variables $\cc^A$ and $\cP_A$ are Grassmann odd ones.
The ghost number grading is introduced by the following
standard prescription:
\begin{equation}
\gh{\cc^A}=1\,,\qquad \gh{\cP_A}=-1\,,
  \label{eq:ghost}
\end{equation}
with all the others variables carrying vanishing ghost
number.

It is natural to consider the ghost variables $\cc$ and $\cP$
as coordinates on the fibres of the respective vector bundles $\Pi\W(\manM)$
and $\Pi\W^*(\manM)$.  Here, $\Pi$ denotes parity
reversing operation; when applied to a vector bundle it transform the
bundle into the super vector bundle with the same base manifold and
the transition functions and the fibres being the Grassmann odd
vector superspaces.

In the BFV-BRST quantization one needs to extend the Poisson structure
on the phase space to the ghost variables, with the variables
$\cc^A$ and $\cP_A$ being canonically conjugated w.r.t.
the bracket.  To this end we consider the following extension of the phase
space $\E_0$:
\begin{equation}
  \label{eq:extended-E}
\E=\mod(\Pi\W(\manM))\oplus \W(\Pi\W(\manM))\,,
\end{equation}
where $\W(\Pi\W(\manM))$ is the vector bundle $\W(\manM)$ pulled back by
the projection $\rho\: \Pi\W(\manM) \to \manM$ and 
$\oplus$ denotes the direct sum of vector bundles over
$\Pi\W(\manM)$.  Note also that construction of modified cotangent
bundle over $\Pi\W(\manM)$ involves the closed 2-form $\rho^*\omega$
defined on $\Pi\W(\manM)$, which is the symplectic form on $\manM$
pulled back by the bundle projection $\rho$.
Identifying $\cc^A$ with the coordinates on the fibres of
$\Pi\W(\manM)$ and $\cP_A$ with their conjugate momenta one can indeed
see that when $\cc^A=\cP_A=0$ extended phase space $\E$ reduces to
$\E_0=\mod\manM\oplus \W(\manM)$.

Symplectic structure~\eqref{eq:symplectic-E0} can be easily extended
to $\E$ by
\begin{equation}
\label{eq:symplectic-E}
\begin{aligned}
\omega^{\,_{\E}}~~=~~~&\pi^*\omega+2dp_i \wedge dx^i+2d\cP_A \wedge d\cc^A+\\
&+~~D_{AB}dY^A\wedge dY^B +
Y^A Y^B d \con_{AB}
-2Y^A\con_{AB} \wedge dY^B\,.
\end{aligned}
\end{equation}
where $\pi^*\omega$ is a 2-form $\omega$ on $\manM$ pulled back by the
projection $\pi\: \E \to \manM$.
The respective Poisson bracket relations are as follows
\begin{equation}
  \begin{split}
    \pb{x^i}{p_j}_{\E}&=\delta^i_j\,, \\
    \pb{Y^A}{Y^B}_{\E}&=D^{AB}\,,\\
    \pb{Y^A}{p_i}_{\E}&=-\con^A_{iB}Y^B\,,\\
    \pb{\cc^A}{\cP_B}_{\E}&=\delta^A_B\,, \\
    \pb{p_i}{p_j}_{\E}&=\omega_{ij}+{\half}R_{ij;\,AB\,}Y^AY^B\,,
  \end{split}
  \label{eq:PB-E}
\end{equation}
with all the others vanishing.  This Poisson bracket can be thought
of as that given by \eqref{eq:extended-PB}, with $\manN=\Pi\W(\manM)$.
This shows that~\eqref{eq:PB-E} determines a globally defined Poisson
bracket on $\E$.

To complete the description of the extended phase space
we specify a class of functions on this space.
Instead of smooth ($\func{\E}$) functions we consider
those which are formal power series in $Y,\cc,\cP$ and
polynomial in $p$ with coefficients in smooth functions
on $\manM$.  The reason is that variables $Y$ serve as the
\textit{conversion variables} and one should allow formal
power series in $Y$.  As for the ghost variables $\cc$ and
$\cP$, each function is always a polynomial in $\cc$ and
$\cP$ since they are Grassmann odd. Let us note, however,
that in the case where $\manM$ is a supermanifold one
should allow formal power series in respective ghost variables.  In
what follows it goes without saying that under the algebra $\cF(\E)$
of ``functions'' on $\E$ we mean the algebra of the power series described
above.  One can check that Poisson bracket~\eqref{eq:PB-E} is well
defined in $\cF(\E)$.  Note also that $\cF(\E)$ can be
considered as the algebra of sections of an appropriate bundle over
$\manM$.

\subsection{Conversion -- classical description}

Now we are in position to proceed with the conversion
of the second-class constraints $\Theta_A$ within the BRST
formalism.  A conversion is to be understood as
the solution of the master equation
\begin{equation}
  \label{eq:master-cl}
  \pb{\Omega}{\Omega}=0\,, \qquad \p{\Omega}=1\,, \quad \gh{\Omega}=1\,,
\end{equation}
(with $\p{\Omega}$ denoting Grassmann parity of $\Omega$)
subjected to the boundary condition
\begin{equation}
\label{eq:boundary-cl}
\Omega|_{Y=0}=\Omega^0=\cc^A\Theta_A=-\cc^ip_i+\cc^\alpha\theta_\alpha\,.
\end{equation}
Let us expand $\Omega$ into the sum of homogeneous
components w.r.t. $Y^A$
\begin{equation}
  \Omega=\sum_{s=0}^\infty \Omega^s\,,
\end{equation}
and assume
that $\Omega^r$ doesn't depend on momenta $p_i$ and ghost
momenta $\cP^A$ for $r \geq 1$.  To construct solution iteratively it is useful to
fix also the first-order term by
\begin{equation}
  \label{eq:1-2-terms}
\Omega^1 = -\,C^A D_{AB}Y^B\,,
\end{equation}
and to introduce a nilpotent operator $\delta$~\cite{[BT],[Fedosov-JDG],[GL]}:
\begin{equation}
  \delta f = \cc^A\dl{Y^A} f\,,\qquad  \delta^2=\delta\delta=0\,.
\end{equation}
If $f$ doesn't depend on the momenta $p_i$ and $\cP_A$ then
\begin{equation}
  \delta f = -\,\pb{\Omega^1}{f}_\E\,,  \qquad f=f(x,Y,\cc)\,.
\end{equation}

An operator $\delta^*$ is introduced by its action on the
homogeneous functions of the form
\begin{equation}
  f_{pq}=f_{A_1,\ldots,A_p\,;\,B_1,\ldots,B_q}(x)\,Y^{A_1}\ldots Y^{A_p}
\cc^{B_1}\ldots \cc^{B_q}
\end{equation}
by means of
\begin{equation}
\begin{split}
  \delta^* f_{pq}&=\frac{1}{p+q}Y^A \dl{\cc^A}f_{pq}\,,  \qquad p+q \neq 0 \\
 \qquad \delta^* f_{00}&=0\,.
\end{split}
\end{equation}
The operator $\delta$ is in some sense inverse to $\delta$
and serve as a contracting homotopy for $\delta$. Indeed,
\begin{equation}
f|_{Y=\cc=0}+\delta\delta^* f + \delta^*\delta  f = f\,.
\end{equation}
\begin{thm}\label{thm:existence-cl}
Given the second-class constraints $\theta_\alpha$ on $\manM$ and
symplectic connection $\con$ in $\W(\manM)$ there exists solution to
the Eq. \eqref{eq:master-cl} satisfying boundary conditions
\eqref{eq:boundary-cl} and \eqref{eq:1-2-terms}.  If in addition one
requires $\delta^*\Omega^r=0$ and $\Omega^r=\Omega^r(x,Y,\cc)$ for $r
\geq 2$ the solution is unique.
\end{thm}
\begin{proof}
In the zeroth order in $Y$ equation~\eqref{eq:master-cl}
implies
\begin{equation}
\begin{split}
\pb{\Omega^0}{\Omega^0}_\E |_{Y=0} +\pb{\Omega^1}{\Omega^1}_\E=0\,.
\end{split}
\end{equation}
This holds provided the boundary conditions~\eqref{eq:boundary-cl} and
\eqref{eq:1-2-terms} are compatible.  In the $r$-th ($r \geq 1$) order
in $Y$
\eqref{eq:master-cl} implies:
\begin{equation}
\label{eq:determin-cl}
\delta\Omega^{r+1}=B^r\,,
\end{equation}
where the quantity $B^r$ is given by
\begin{equation}
  \label{eq:B-explicit}
  \begin{split}
B^1&=\pb{\Omega^0}{\Omega^1}_\E\,,\\[3pt]
B^2&=\pb{\Omega^0}{\Omega^2}_\E
+\half \pb{\Omega^2}{\Omega^2}_\E+
\frac{1}{4}\cc^i \cc^j  {\bar R}_{ij;\, A B} Y^A Y^B\,,\\
B^r&=\pb{\Omega^0}{\Omega^r}_\E+\half\sum_{s=0}^{r-2}
\pb{\Omega^{s+2}}{\Omega^{r-s}}_\E\,,\quad r\geq 3\,,
\end{split}
\end{equation}
and we have assumed that $\Omega^r=\Omega^r(x,Y,\cc)$ for $r\geq 2$.
The necessary and sufficient condition for Eq. \eqref{eq:determin-cl}
to have a solution is $\delta B^r=0$. Let us first show explicitly that
$\delta B^1=0$. Indeed, in view of the zeroth order equation
\begin{equation}
  \begin{split}
\delta B^1=\delta\pb{\Omega^0}{\Omega^1}_\E=-\pb{\Omega^1}{\pb{\Omega^1}{\Omega^0}_\E}_E=
{-\, \frac{1}{2}}\pb{\pb{\Omega^1}{\Omega^1}_\E}{\Omega^0}_\E=\\
=\half\pb{\pb{\Omega^0}{\Omega^0}_\E{\bigr|}_{Y=0}}{\Omega^0}_\E=0\,.
\end{split}
\end{equation}
Then a particular solution for $\Omega^2$ is
\begin{equation}
  \Omega^2=\delta^*\pb{\Omega^0}{\Omega^1}\,.
\end{equation}

The proof of the statement goes further along the standard
induction procedure~\cite{[BT],[GL]}: one can first check that
$\delta B^s=0$ provided $\Omega^r$ satisfy \eqref{eq:determin-cl}
for $r\leq s-1$; one then finds:
\begin{equation}
\Omega^{s+1}=\delta^*B^s\,.
\end{equation}
Finally, one can check that $\Omega^{s+1}=\delta^*B^s$ is a unique
solution of Eq.~\eqref{eq:determin-cl} for $r=s$ provided the
additional condition $\delta^* \Omega^{s+1}=0$ is imposed.
\end{proof}
It follows from the Theorem~\bref{thm:existence-cl} that we
have arrived at the first-class constraint theory whose extended phase
space is $\E$. Since under the additional condition
$\delta^*\Omega^r=0\,,\, r \geq 2$ classical BRST charge is unique and
is obviously linear in $\cc$, this first class constraint system is an
Abelian one.

\subsection{BRST cohomology}
By definition, an observable of the
BFV-BRST system determined by the BRST charge
$\Omega$ is a function $f$ on the extended phase
space satisfying
\begin{equation}
\pb{\Omega}{f}_\E=0\,, \qquad \gh{f}=0\,.
\end{equation}
Two observables $f$ and $g$ are said equivalent iff their difference
is BRST exact, i.e. $f-g=\pb{\Omega}{h}_\E=0$ for some function
$h\in\cF(\E)$. The space of inequivalent observables of the system is
thus the BRST cohomology with ghost number zero (i.e. quotient of the
BRST closed functions with zero ghost number modulo exact ones).  Let
us now investigate the structure of the BRST cohomology of the BFV
system on $\E$ determined by the BRST charge $\Omega$ constructed in
\bref{thm:existence-cl}.
\begin{prop}\label{prop:extension-cl}
\begin{enumerate}
\item 
Let $f_0$ be an arbitrary $Y$-independent function (i.e. a function of
$x,p,\cc,\cP$ only) of a nonnegative ghost number. Then there exist
function $f$ on $\E$ such that
\begin{equation}
\label{eq:c-extension}
\pb{\Omega}{f}_{\E}=0\,, \quad f{\bigr|}_{Y=0}=f_0\,,\qquad \gh{f}=\gh{f_0}\,.
\end{equation}
Moreover, if $\tilde f$ also satisfies \eqref{eq:c-extension} then
\begin{equation}
  \label{eq:c-arbitrarines}
{\tilde f} -f =\pb{\Omega}{g}_\E
\end{equation}
for some function $g$ on $\E$.
\item If in addition one requires $f$ to satisfy $\delta^* (f-f_0)=0$
then $f$ is the unique solution to~\eqref{eq:c-extension}.
\end{enumerate}
\end{prop}
\begin{proof}
Let us represent adjoint action $\pb{\Omega}{\cdot}_\E$ of the BRST
charge and function $f$ as the sum of homogeneous components
w.r.t. $Y$:
\begin{equation}
\pb{\Omega}{\cdot}_\E=-\,\delta+\sum_{s=0}^\infty \delta_s\,, \qquad
f=\sum_{s=0}^\infty f_s\,,
\end{equation}
with $\delta=\cc^A\dl{Y^A}$.  In the $r$-th order in $Y$
equation \eqref{eq:c-extension} then becomes
\begin{equation}
\label{eq:determin-cl2}
\delta f_{r+1}=B_r\,,
\end{equation}
where $B_r$ is given by
\begin{equation}
  B_r=\sum_{s=0}^r \delta_s f_{r-s}\,.
\end{equation}
The consistency condition for equation~\eqref{eq:determin-cl2} is
$\delta B_r=0$.  Let us show that $\delta B_p=0$ provided
\eqref{eq:determin-cl2} is fulfilled for any $r\leq p$ and $f_0$
carries nonnegative ghost number.  Indeed, in the zeroth order in $Y$
\eqref{eq:determin-cl2} rewrites as
\begin{equation}
  \delta f_1=\delta_0 f_0\,.
\end{equation}
The consistency condition is obviously fulfilled since
\begin{equation}
  \delta B_0=\delta \delta_0 f_0=-\delta_0 \delta f_0=0\,.
\end{equation}
The later equality follows because $f_0$ is independent of $Y$.
Assume that $f_r$ are given for $r \leq p$ and
equation \eqref{eq:determin-cl2} is fulfiled for $r \leq p-1$.
Consider then the identity:
\begin{equation}
\label{eq:identity}
\pb{\Omega}{\pb{\Omega}{\sum_{s=0}^p f_s}_\E}_\E =
(-\,\delta+\sum_{q=0}^\infty \delta_q)
(-\,\delta+\sum_{t=0}^\infty \delta_t)\sum_{s=0}^p f_s=0\,.
\end{equation}
One can see that
\begin{equation}
(-\,\delta+\sum_{t=0}^\infty \delta_t)\sum_{s=0}^p f_s=B_p+\cdots
\end{equation}
were $\ldots$ denote terms of order higher than $p$.
In the $p-1$-th order in $Y$ Eq. \eqref{eq:identity}
then implies: $\delta B_p=0$.

That $\delta B_p=0$ allows one to construct
solution iteratively:
\begin{equation}
f_{p+1}=\delta^*B_p\,.
\end{equation}
One can indeed check that
\begin{equation}
\delta f_{p+1}=\delta \delta^* B_p=
\delta \delta^* B_p+\delta^* \delta B_p +B_p{\bigr|}_{\cc=Y=0}=B_p\,,
\end{equation}
since $\delta B_p=0$ and $B_p{\bigr|}_{\cc=Y=0}=0$.  The later
equality is obvious for $p \geq 1$; the fact that
$B_0{\bigr|}_{\cc=Y=0}=(\delta_0 f_0){\bigr|}_{\cc=Y=0}=0$
follows because $f_0$ has a nonnegative ghost number.
Thus the first part of the statement is proved.

Further, let $f$ and $\tilde f$ satisfy~\eqref{eq:c-extension}.
Assume that $f_s={\tilde  f}_s$ for any $s \leq r$.  In the
$r$-th order in $Y$ Eq.~\eqref{eq:c-extension} then implies:
\begin{equation}
\delta f_{r+1}=B_r\,, \qquad \delta {{\tilde f}_{r+1}}=B_r\,.
\end{equation}
Since $\delta({\tilde f}_r-f_r)=0$ and
$({\tilde f}_r-f_r)\bigr|_{Y=\cc=0}=0$,
one can represent ${\tilde f}_r-f_r$ as
\begin{equation}
  {\tilde f}_r-f_r=\delta g_r\,, \qquad g_r=\delta^*({\tilde f}_r-f_r)\,.
\end{equation}
Thus
\begin{equation}
{\bar f}=f+\pb{\Omega}{g_r}_\E
\end{equation}
satisfies \eqref{eq:c-extension} and coincides with $\tilde f$ up to
to the terms of order higher than $r+1$. Iteratively applying
this procedure one can construct function $g$
such that
\begin{equation}
  {\tilde f}=f+\pb{\Omega}{g}_\E\,.
\end{equation}

Finally, let $f$ and $\tilde f$ satisfy~\eqref{eq:c-extension}
and additional condition
$\delta^* (f-f_0)=\delta^* ({\tilde f}-f_0)=0$. For
$d_{r+1}=f_{r+1}-{\tilde f}_{r+1}$ one then has:
$$
\delta d_{r+1}=\delta^* d_{r+1}=0\,.
$$
This implies that $d_{r+1}=0$ because $d_{r+1}$ is at least linear in
$Y$.  This proves second item.
\end{proof}
\begin{lemma}\label{lemma:koszul-cl}
Let $f_0$ be an arbitrary $Y$-independent function of
nonnegative  ghost number.  Let also $f$ be a BRST invariant
extension of $f_0$ obtained by Proposition~\bref{prop:extension-cl}.  Then
\begin{equation}
  f=\pb{\Omega}{h}_\E\,,
\end{equation}
for some function $h$ on $\E$ if and only if
\begin{equation}
  f_0{\bigr|}_{\cc^A=\theta_\alpha=p_i=0}=0\,.
\end{equation}
\end{lemma}
\begin{proof}
It is useful to introduce new coordinate functions
\begin{equation}
 {\bar p}_i=p_i-\cP_k\Gamma^k_{ij}\cc^j\,,
\end{equation}
where $\Gamma^k_{ij}$ are the coefficients of an arbitrary symmetric
connection on $\manM$.  The reason is that unlike $p_i$ that have
inhomogeneous transformation properties, the coordinate functions
${\bar p_i}$ transform as the coefficients of a 1-form on $\manM$.
Note, that functions ${\bar p},x,Y,\cc$ and $\cP$ also
form a local coordinate system on $\E$ and conditions
$\cc^A=\theta_\alpha=p_i=0$ and $\cc^A=\theta_\alpha={\bar p}_i=0$ are
obviously equivalent.

Any BRST exact function $f$ (i.e. a function that can be represented
as $f=\pb{\Omega}{h}_\E$) evidently vanishes when
$Y^A=\cc^A=\theta_\alpha={\bar p}_i=0$. Conversely, assume that $f_0$
vanishes when $\cc^A=\theta_\alpha={\bar p}_i=0$ and carries
nonnegative ghost number. Then it can be represented as
\begin{equation}
f_0=\cc^A f_A+\theta_\alpha f^\alpha + {\bar p}_i f^i \,,
\end{equation}
where functions $f^i$ and $f_\alpha$ can be taken in the form $f^i=f^i(x,\bar p)$
and $f_\alpha=f_\alpha(x)$ respectively.
One can also choose $f_A$ and $f^i$ such that they transform
as the components of a section of $\W^*(\manM)$ and
components of a vector field on $\manM$ respectively.
We introduce
\begin{equation}
\begin{split}
{\bar f}^\alpha&=f^\alpha+Y^l \d_l f^\alpha\,,\\
{\bar f}^i&= f^i +
Y^l\, (\d_l f^i + \Gamma^i_{lk} f^k - {\bar p}_j \Gamma^j_{lk} \dl{{\bar p}_k}
f^i + \half \cc^j R^m_{ljk}\cP_m \dl{{\bar p}_k}f^i)\,,
\end{split}
\end{equation}
where $R$ is the curvature of $\Gamma$.  Note
that ${\bar  f}^i$ transform as components of a vector field on
$\manM$; in particular, ${\bar f}^i\cP_i$ is the globally defined
function on $\E$.  Picking $h_0$ as
\begin{equation}
{h_0}=
-Y^A f_A
+\cP_\alpha {\bar f}^\alpha
-\cP_i {\bar f}^i
 \,,
\end{equation}
one can indeed check that
\begin{equation}
f_0=\left( \pb{\Omega}{{h_0}}_\E \right)|_{Y=0}\,.
\end{equation}
Finally, it follows from Proposition~\bref{prop:extension-cl} that there
exists function $h_1$ such that $f=\pb{\Omega}{h_0+h_1}_\E$.
\end{proof}

As a consequence of Proposition~\bref{prop:extension-cl} and
Lemma~\bref{lemma:koszul-cl} we arrive at the following Theorem.
\begin{thm}\label{thm:H-cl}
In the nonnegative ghost number the BRST cohomology of the BRST charge
$\Omega$ constructed in Theorem \bref{thm:existence-cl} is:
\begin{equation}
\begin{split}
H^n=0\,,&\quad n\geq 1\\
H^n=\func{\Sigma}\,, & \quad n=0\,.
\end{split}
\end{equation}
where $H^n$ denotes cohomology with ghost number $n$ and
$\func\Sigma$ is an algebra of smooth functions on the
constraint surface $\Sigma \subset \manM$.
\end{thm}
It follows from the Theorem~\bref{thm:H-cl} that at least at the classical level,
the original second-class constraint system on $\manM$ is equivalent
to the constructed BFV-BRST system on $\E$.

\section{Quantum description and star-product}\label{sec:q}

In this section we quantize the constructed BFV-BRST system.
This includes constructing quantum BRST charge, quantum BRST
observables and evaluating quantum BRST cohomology. The star product
for the Dirac bracket on $\manM$ is constructed as the quantum
multiplication of BRST observables.

\subsection{Quantization of the extended phase space}\label{subsec:quantization}
The extended phase space $\E$ of the BFV-BRST system is in general
a non-flat manifold and thus can not be quantized directly.
Fortunately, all physical observables
as well as the generators of the BRST algebra
(the BRST charge $\Omega$ and the ghost charge $G=\cc^A\cP_A$)
can be chosen as elements of a certain subalgebra
$\aA\subset\cF(\E)$.  In its turn $\aA$ is closed
w.r.t. Poisson bracket and can be explicitly quantized.  This implies
that one can equip $\aA$ with the quantum multiplication satisfying
the standard correspondence principle.

Since the construction of $\aA$ is a direct generalization
of that from~\cite{[GL]} we present it very brief here.
Let $\aA_0$ be a subalgebra of functions on $\E$,
which do not depend on the momenta $p_i$ and the ghost momenta
$\cP_i$.  A general element of this algebra is then given
by
\begin{equation}
  a=a(x^i,Y^A,\cc^A,\cP_\alpha)\,.
\end{equation}
In invariant terms $\aA_0$ is a tensor product
of algebra generated by $\cP_\alpha$ and algebra of functions
on $\Pi W(\manM)\oplus W(\manM)$ pulled back
by the projection $\E \to \Pi W(\manM)\oplus W(\manM)$. $\aA_0$ is a Poisson algebra,
i.e. it is closed w.r.t. the ordinary multiplication and the Poisson
bracket.

At the quantum level it is useful to consider the algebra
\begin{equation}
 \qA_0\equiv\aA_0\tensor [[\hbar]]\,,
\end{equation}
where $[[\hbar]]$ denotes the algebra of formal power series in
$\hbar$. $\qA_0$ is also a Poisson algebra.  It is easy to obtain
a deformation quantization of $\qA_0$ considered as the Poisson
algebra.  Indeed the Weyl multiplication works well.  Namely, for
any $a,b\in \qA_0$ one postulates
\begin{multline}
\label{eq:Weyl-star-product}
(a \star b)(x,Y,\cc,\cP,\hbar)~~=\\
=~~\{(a(x,Y_1,\cc_1,\cP_2,\hbar) exp ( -\frac{i\hbar}{2} (D^{AB}
\dr{Y_1^A} \dl{Y_2^B}+\dr{\cc_1^\alpha}\dl{\cP^2_\alpha}+
\dr{\cP^1_\alpha}\dl{\cc_2^\alpha}))\\
b(x,Y_2,\cc_2,\cP_2,\hbar)\}
{\bigr|}_{Y_1=Y_2=Y,\,\cc_1=\cc_2=\cc,\,\cP_1=\cP_2=\cP} \,\,,
\end{multline}
where $\cP$ stands for the dependence on $\cP_\alpha$ only.

Subalgebra $\qA$ ($\aA$) is an extension of
$\qA_0$ (respectively $\aA_0$) by
elements $\P=-\cc^i p_i,\,\G=\cc^i \cP_i$.  A general homogeneous
element $a \in \aA$ is then given by
\begin{equation}
a=\P^r \G^s a_0\,, \quad r=0,1\,,
\quad s=0,1,\ldots, 2N\,,
\quad a_0\in \qA_0\,.
\end{equation}
Weyl product~\eqref{eq:Weyl-star-product} can be easily
extended from $\qA_0$ to $\qA$.  Explicit construction of the
quantum multiplication in $\qA$ is an obvious generalization
of that presented in~\cite{[GL]}.  Here we write explicitly only the
multiplication table for $\G$-independent elements (it turns out that
it is sufficient for present considerations):
\begin{equation}
\label{eq:P-star}
\P\star a=\P a\,,\qquad
a\star \P=a\P + (-1)^{\p{a}}\frac{i\hbar}{2}\,\diff-con a \,, \qquad
 \P\star\P=\ih( {\bar R}+\omega)\,,
\end{equation}
where $a$ is an arbitrary element of $\qA_0$,
$\diff-con=\cc^i\d_i-\cc^i\con^A_{iB}Y^B\dl{Y^A}\,,$ and
functions $\bar R$ and $\omega$ are the ``generating functions'' for the
curvature and symplectic form:
\begin{equation}
\label{eq:R-D-omega}
{\bar R}=\frac{1}{4} \cc^i \cc^j  {\bar R}_{ij;\, A B} Y^A Y^B \,,
\qquad
\omega=\half \cc^i \omega_{ij}\cc^j\,.\qquad {\bar R}\,,\omega\in\aA\subset\qA\,.
\end{equation}
\maxim{Sdes` ya by zamenil $\omega\in\aA\subset\qA$ na $\omega\in\hat\aA_0\subset\qA$}

In what follows we treat $\qA$ as an associative algebra with the 
product determined by~\eqref{eq:Weyl-star-product} and \eqref{eq:P-star}
and extended to $\G$-dependent elements as in~\cite{[GL]}.  Let us
introduce a useful grading in $\qA$.  Namely,
we prescribe the following degrees to the variables:
\begin{equation}
\begin{split}
  \label{eq:degrees}
  \deg{x^i}=\deg{\cc^A}=0\,,&\qquad \deg{p_i}=\deg{\cP_A}=2 \,,\\
  \deg{Y^A}=1\,,& \qquad  \deg{\hbar}=2\,.
\end{split}
\end{equation}
The quantum multiplication in $\qA_0$ obviously preserves the degree.

\subsection{Quantum BRST charge}
Since the classical BRST charge from Theorem~\bref{thm:existence-cl}
belongs to $\qA$ it is natural to a define quantum BRST
charge $\hat\Omega$ as a solution to the quantum master equation
\begin{equation} \label{eq:master-q}
\qcommut{{\hat\Omega}}{{\hat\Omega}} \equiv
2\,{\hat\Omega}\star {\hat\Omega}=0\,, 
\qquad {\hat\Omega} \in \qA\,,\qquad \p{\hat\Omega}=1\,,\,\, \gh{\hat\Omega}=1\,.
\end{equation}
In analysis of the master equation it is useful to expand $\hat\Omega$
into the sum of homogeneous components w.r.t. degree introduced
in~\eqref{eq:degrees}:
\begin{equation}
  \label{eq:Omega-expansion}
  {\hat\Omega}=\sum^\infty_{r=0} {\hat\Omega}^r\,,\qquad \deg{{\hat\Omega^r}}=r\,.
\end{equation}
The appropriate quantum counterparts of the classical boundary
conditions~\eqref{eq:boundary-cl} and~\eqref{eq:1-2-terms}
chosen in~\bref{thm:existence-cl} are given by:
\begin{equation}
\label{eq:boundary}
  {\hat\Omega}^0=\cc^\alpha \theta_\alpha\,,\qquad
  {\hat\Omega}^1=-\cc^A D_{AB}Y^B\,,\qquad
  {\hat\Omega}^2=-\cc^i p_i+\frac{i}{\hbar}\delta^*(\qcommut{\hat\Omega^0}{\hat\Omega^1})\,.
\end{equation}
\begin{thm}\label{thm:existence-q}
Equation~\eqref{eq:master-q} has a solution $\hat\Omega$ satisfying
boundary condition~\eqref{eq:boundary}. If in addition
one requires $\hat\Omega$ to satisfy $\delta^*{\hat\Omega}^r=0$
and ${\hat\Omega}^r={\hat\Omega}^r(x,Y,\cc,\hbar)$ for $r\geq 3$ the solution is
unique.
\end{thm}
\begin{proof}
The proof is a direct generalization of that of the analogous
statement from~\cite{[GL]}. Assume that ${\hat\Omega}^r\in \qA_0$ for $r\geq 3$
and ${\hat\Omega}$ doesn't depend on $\cP$. Then the solution can be
constructed iteratively in the form
$$
{\hat\Omega}^{r+1}=\delta^*{\hat B}^r\,,\qquad r\geq 2\,,
$$
where $\deg{{\hat\Omega}^{a}}=a$ and ${\hat B}^r$ is given by
\begin{equation}
\label{eq:B-explicit-q}
{\hat B}^r=\frac{i}{2\hbar}\sum_{s=0}^{r-2}
\qcommut{{\hat\Omega}^{s+2}}{{\hat\Omega}^{r-s}}\,,\qquad r\geq 2\,.
\end{equation}
\end{proof}

\subsection{Quantum BRST observables and star-product}
At the classical level each physical observable (an element of the zero ghost
number BRST cohomology) can be considered as an element of $\qA$.
It is then natural to define a quantum BRST observable
as a function $f$ satisfying
\begin{equation}
\label{eq:extension-q}
  \qcommut{\hat\Omega}{f}=0\,, \qquad f\in \qA\,.
\end{equation}
Two observables are said equivalent iff their difference
can be represented as $\frac{i}{\hbar}\qcommut{\hat\Omega}{g}$
for some $g\in \qA$. Inequivalent observables is thus
a zero ghost number cohomology of
$\frac{i}{\hbar}\qcommut{\hat\Omega}{\cdot}$.

Let us consider first observables in $\qA_0$.  It turns out that any
function $f_0(x,\cc)$ admits a BRST invariant extension $f$
satisfying \eqref{eq:extension-q} and
the boundary condition
\begin{equation}
\label{eq:boundary-q-ext}
  f{\bigr|}_{Y=0}=f_0\,.
\end{equation}
If $f_0$ has a definite ghost number we also
require: $\gh{f}=\gh{f_0}$.
\begin{prop}\label{prop:extension-q}
Given a function $f_0(x,\cc)$ there exists solution
$f\in \qA_0$ to the equation \eqref{eq:extension-q} satisfying
boundary condition~\eqref{eq:boundary-q-ext}.  If in
addition one requires $f$ to satisfy $\delta^*(f-f_0)=0$
and $f=f(x,Y,\cc,\hbar)$ then the solution is unique.
\end{prop}
\begin{proof}
The proof is a direct generalization of that of the analogous
statement from~\cite{[GL]}. Let us expand $f$ as
\begin{equation}
f=f_0+\sum_{s=1}^\infty f_s\,,\qquad \deg{f_s}=s\,.
\end{equation}
The solution is constructed
iteratively in the form
$$
{f}_{r+1}=\delta^*{B}_r\,,
$$
with $B^r$ being
$$
B_r=\frac{i}{2\hbar}\sum_{s=0}^{r-2}
\qcommut{{\hat\Omega}^{s+2}}{f_{r-s}}\,.
$$
In particular, for the function $f_0=f_0(x)$ we
have explicitly
\begin{equation}
  f=f_0+Y^i \d_i f_0 +\cdots\,,
\end{equation}
where $\cdots$ denotes terms of higher order in $Y$ and $\hbar$.
\end{proof}
Because the BRST invariant extension determined by the statement is
obviously a linear map it can be extended to functions
depending formally on $\hbar$.

By means of Proposition~\bref{prop:extension-q} statement we obtain a
one-to-one correspondence between $\func{\manM}\tensor[[\hbar]]$ and
the BRST invariant functions depending on $x,Y$ and $\hbar$ only. The
space of these functions is obviously closed w.r.t. the quantum
multiplication~\eqref{eq:Weyl-star-product} in $\qA_0$. This
multiplication determines thus a star product on $\manM$, giving a
deformation quantization of the Dirac bracket on $\manM$.  Namely, given
functions $f_0$ and $g_0$ on $\manM$ one has
\begin{equation}
  f_0\star_{\manM} g_0=(f \star g){\bigr|}_{Y=0}=
f_0 g_0-\frac{i\hbar}{2}\pb{f_0}{g_0}^D_{\manM}+\cdots\,,
\end{equation}
where $f=f(x,Y,\hbar)$ and $g=g(x,Y,\hbar)$ are the unique quantum
BRST invariant extensions of $f_0$ and $g_0$ obtained by
Proposition~\bref{prop:extension-q}, $\star$ is a Weyl product given
by~\eqref{eq:Weyl-star-product}, and $\cdots$ denote terms of higher
order in $\hbar$.

\subsection{BRST cohomology at the quantum level}\label{subsec:BRST-q}
To complete description of the constructed quantum gauge system
let us calculate cohomology of the quantum BRST charge obtained in
Theorem~\bref{thm:existence-q}.  Since we do not have
a quantum multiplication in the algebra $\cF(\E)$ of functions
on the entire $\E$ but only in the subalgebra $\qA \subset \cF(\E)$
we are interested in the cohomology of $\hat\Omega$ evaluated in
$\qA$.

Let us first note that cohomology of the classical BRST charge
$\Omega$ evaluated in $\qA$ (considered as a Poisson algebra)
coincides with that calculated in $\cF(\E)\tensor[[\hbar]]$. Indeed,
appropriate modifications of Proposition~\bref{prop:extension-cl} and
Lemma~\bref{lemma:koszul-cl} show this.

Instead of formulating quantum counterparts of
Proposition~\bref{prop:extension-cl} and Lemma~\bref{lemma:koszul-cl}
we construct the quantum BRST invariant elements as the quantum
deformations of the respective classical ones.
\begin{prop}\label{prop:hbar-extension}
\begin{enumerate}
\item
Let $f\in\aA$ has a nonnegative ghost number
and satisfies
\begin{equation}
  \pb{{\hat\Omega}\bigr|_{\hbar=0}}{f}_\E=0\,,
\end{equation}
where $\hat\Omega$ is the quantum BRST charge obtained in
Theorem~\bref{thm:existence-q}.  Then there exist the quantum
corrections $f_s\,,s\geq 1$ such that $f_s\in \qA,\, \gh{f_s}=\gh{f},$
and ${\hat f}=f\ih f_1+(\ih)^2 f_2+\ldots$ satisfies
\begin{equation}
\label{eq:deform}
  \qcommut{\hat\Omega}{\hat f}=0\,.
\end{equation}
\item
Let $\hat f\in \qA$ carries a nonnegative ghost number
and satisfies~\eqref{eq:deform}. Then there exists $\hat g \in \qA$
such that ${\hat f}+\frac{i}{\hbar}\qcommut{\hat\Omega}{\hat g}$ is
the function of $x,Y,\cc$ and $\hbar$ only.
\end{enumerate}
\end{prop}
\begin{proof}
Function $f_0$ is a classical BRST observable, because of
$\Omega\equiv{\hat\Omega}\bigr|_{\hbar=0}$ is obviously a classical
BRST charge from~\bref{thm:existence-cl}.  In the $r+2$-th order in
$\hbar$ Eq.~\eqref{eq:deform} implies
\begin{equation}
\label{eq:hbar-determin}
  \pb{\Omega}{f_{r+1}}_\E+\cB_r=0\,,
\end{equation}
with $\cB_r$ given by
\begin{equation}
\cB_r=\sum_{s+t+u=r+2} \qcommut{{\hat\Omega}^s}{f_t}^u\,, \quad s,u\geq
1\,, \quad t\geq 0\,,
\end{equation}
where we have expanded ${\hat\Omega}$ and $\qcommut{\,}{}$ in $\hbar$
according to
\begin{equation}
{\hat\Omega}=\Omega+\sum^\infty_{s=1}(\ih)^s {\hat\Omega}^s\,,\qquad
\qcommut{\cdot\,}{\cdot}=\sum^\infty_{s=1}(\ih)^s \,
\qcommut{\cdot\,}{\cdot}^s\,.
\end{equation}
Arguments similar to those of the proof for
Proposition~\bref{prop:extension-cl} show that
$\pb{\Omega}{\cB^p}_\E=0$ provided Eq.~\eqref{eq:hbar-determin} holds
for all $r\leq p-1$. Since cohomology of a classical BRST charge
vanish in strictly positive ghost number and $\gh{\cB_{r+1}}\geq 1$ we
see that Eq.~\eqref{eq:hbar-determin} has a solution for $r=p$.
Moreover, $f_{r+1}$ can be taken to belong to $\qA$.

The second part of the statement is trivial
when $\hat f$ carries strictly positive ghost number,
since classical BRST cohomology vanishes in this case.
Let ${\hat f}\in\qA$ satisfies~\eqref{eq:deform}
and  $\gh{\hat f}=0$. Let us show that there exists $\hat g$
such that ${\hat f}+\frac{i}{\hbar}\qcommut{\hat\Omega}{\hat g}$
depends on $x,Y$ and $\hbar$ only. This is obvious for the zero order
term $f$ in the expansion ${\hat f}=f\ih f_1+(\ih)^2 f_2+\ldots$
of $\hat f$ in powers of $\hbar$.  Assume that this is the case
for all $f_s,\,s\leq r$. In the $r+2$-th order in $\hbar$
Eq.~\eqref{eq:deform}
then takes the form \eqref{eq:hbar-determin}. Because $f_s$ is a function
of only $x,Y$ for $s\leq r$ then $\cB_r$ in \eqref{eq:hbar-determin}
has the form $\cc^A b_A(x,Y)$; one can then find
${\tilde  f}_{r+1}(x,Y)$ such that
\begin{equation}
  \pb{\Omega}{{\tilde f}_{r+1}}_\E+\cB_r=0\,.
\end{equation}
Since $\pb{\Omega}{f_{r+1}-{\tilde f}_{r+1}}_\E=0$
there exist functions ${\bar f}_{r+1}(x,Y)$ and $g_{r+1}\in\qA$ such that
\begin{equation}
  f_{r+1}={\bar f}_{r+1}-\pb{\Omega}{g_{r+1}}_\E\,.
\end{equation}
It follows that ${\bar f}_{r+1}(x,Y)$
is a $r+1$-th order term in the expansion of
${\hat f}+\frac{i}{\hbar}\qcommut{\hat\Omega}{(\ih)^{r+1}g_{r+1}}$
in powers of $\hbar$. Proceeding further
by induction one can find $\hat g$ required in item $2$.
\end{proof}
Proposition~\bref{prop:extension-q}
establishes a one-to-one correspondence between $\func{\manM}\tensor
[[\hbar]]$ and the quantum BRST invariant functions depending
on $x,Y$ and $\hbar$ only. Among these functions those vanishing
when $Y=\theta=0$ are BRST trivial (this can be checked directly
or by means of the arguments similar to those in the proof of
item 2 of Proposition above) while those which do not
vanish when $Y=\theta=0$ are obviously nontrivial. In this way we
arrive at the quantum counterpart of Theorem~\bref{thm:H-cl}:
\begin{thm}\label{thm:H-q}
In nonnegative ghost number the quantum BRST cohomology
of $\hat\Omega$ evaluated in $\qA$ is given by
\begin{equation}
\begin{split}
{\hat H}^n=0\,,&\quad n\geq 1\\
{\hat H}^n=\func{\Sigma}\tensor[[\hbar]]\,, & \quad n=0\,.
\end{split}
\end{equation}
\end{thm}
It follows from the Theorem~\bref{thm:H-q} that the constructed gauge
system is equivalent to the original second-class system
on $\manM$ at the quantum level as well. This, in particular,
implies that the quantum multiplication of inequivalent quantum BRST
observables determines a star-product on $\Sigma$.

Indeed, the quantum multiplication in $\aA_0$ determines a
quantum multiplication in ${\hat H}^{0}$, which,
in turn, is isomorphic to $\func{\Sigma}$.
The isomorphism $\func{\Sigma}\tensor[[\hbar]] \to {\hat H}^0$
as well as the inverse map can be obtained by
Proposition~\bref{prop:extension-q}.  Namely, each function
$f_\Sigma \in \func{\Sigma}\tensor[[\hbar]]$ can be represented as
$f_0|_{\Sigma}$ for some function
$f_0\in\func{\manM}\tensor[[\hbar]]$. Thus the isomorphism map
$\func{\Sigma}\tensor[[\hbar]]\to {\hat H}^0$
is given explicitly by the quantum BRST invariant
extension of $f_0$ determined by Proposition~\bref{prop:extension-q}.
The inverse map is obviously given by the restriction to $\Sigma$.
The star-product on $\Sigma$ is then given by
\begin{equation}
\label{eq:star-Sigma}
f_\Sigma \star_\Sigma g_\Sigma = (f_0 \star_\manM g_0){\bigr|}_\Sigma=
({f}\star{g}){\bigr|}_{\Sigma} =
f_\Sigma g_\Sigma-\frac{i}{2\hbar}\pb{f_\Sigma}{g_\Sigma}_{\Sigma}+\cdots\,,\qquad
f_0|_{\Sigma}=f_\Sigma,\,\,  g_0|_{\Sigma}=g_\Sigma\,,
\end{equation}
where $f$ and $g$ are the quantum BRST invariant
extensions of functions $f_0$ and $g_0$ respectively,
$\pb{\cdot}{\cdot}_\Sigma$ denotes the Poisson bracket on $\Sigma$
(restriction of the Dirac bracket $\pb{\cdot}{\cdot}^D_\manM$ to $\Sigma$),
and $\cdots$ denote higher order terms in $\hbar$.
It follows that the star-product~\eqref{eq:star-Sigma} is well defined
in the sense that it doesn't depend on the choice of
representatives
$f_0,g_0\:f_0|_{\Sigma}=f_\Sigma\,,\,g_0|_{\Sigma}=g_\Sigma$
of functions $f_\Sigma,g_\Sigma$ on $\Sigma$. In fact, it also doesn't
depend on the choice of a BRST invariant extensions for $f_0$ and $g_0$.

\subsection{Quantization with the Dirac connection and the center of
  the star-commutator algebra}\label{subsec:special}
Let us make some general observation on the structure of the BRST
charge and the BRST invariant observables in the case where the
symplectic connection $\con$ entering the symplectic structure
of $\E$ is the specific connection $\con^0$ given by~\eqref{eq:con0-explicit}.
Although the proposed quantization scheme works well with an
arbitrary symplectic connection in $\W(\manM)$, it
turns out that when one uses the specific connection $\con^0$,
the central functions of Dirac bracket are also central functions of
the respective $*$-commutator on $\manM$.

Let us
consider first the structure of the quantum BRST charge
$\hat\Omega$.
\begin{prop}\label{prop:Omega-special}
Let the symplectic connection entering the symplectic structure
be the specific connection $\con^0$ given by~\eqref{eq:con0-explicit}.
Let also ${\hat\Omega}_0$ be the unique solution to the master
equation~\eqref{eq:master-q} obtained by
Theorem~\bref{thm:existence-q} with the boundary
conditions~\eqref{eq:boundary} and the additional conditions
$\delta^*{\hat\Omega}_0^r=0$ and
${\hat\Omega}_0^r={\hat\Omega}_0^r(x,Y,\cc,\hbar)$ for $r \geq 3$.
Then ${\hat\Omega}_0$ satisfies
\begin{equation}
\label{eq:property}
  \dl{\cc^\alpha}{\hat\Omega}_0^r=\dl{Y^\alpha}{\hat\Omega}_0^r=0\,,
\qquad r \geq 3\,, \qquad \deg{{\hat\Omega}^r_0}=r\,.
\end{equation}
The classical BRST charge $\Omega_0$ can be obtained as
$\Omega_0=({\hat\Omega}_0){\bigr|}_{\hbar=0}$ and is then also
$Y^\alpha$ and $\cc^\alpha$ independent (except the terms of zero and
first order in $Y$).
\end{prop}
\begin{proof}
For the degree $2$ term one has ${\hat\Omega_0}^2=-\cc^i p_i$,
since $\qcommut{{\hat\Omega}_0^0}{{\hat\Omega}_0^1}$ vanishes.
Assume that~\eqref{eq:property} holds for $r\leq s$.  Then
${\hat\Omega}_0^{s+1}$ is also $\cc^\alpha$ and
$Y^\alpha$ independent.  Indeed, for ${\hat\Omega}_0^{s+1}$ one has
\begin{equation}
{\hat\Omega}_0^{s+1}=\delta^*{\hat B}^s\,, 
\end{equation}
with ${\hat B}^s$ given by~\eqref{eq:B-explicit-q}. 
The space of $\cc^\alpha$ and $Y^\alpha$ independent functions is
closed w.r.t. the quantum multiplication in $\qA$, since the respective
components of the connection $\con^0$ and its curvature vanish.
In its turn operator $\delta^*$ also preserves this space. Thus ${\hat\Omega}_0^{s+1}$
satisfies~\eqref{eq:property} and the statement follows by
induction.  Up to the terms of degree higher than three
$\hat\Omega_0$ has the following form
\begin{equation}
  \label{eq:3}
{\hat\Omega}_0=\cc^\alpha \theta_\alpha - \cc^i p_i-\cc^A D_{AB}Y^B
-\frac{1}{8} \cc^i {\bar R}^0_{ij;\,kl}Y^jY^kY^l+\ldots\,.
\end{equation}
\end{proof}

The same concerns the BRST invariant extension of functions from
$\manM$ both quantum and classical, which is obtained w.r.t. quantum
BRST charge ${\hat\Omega}_0$ and classical BRST charge $\Omega_0$
respectively.
\begin{prop}
\label{prop:extension-special}
Let $f_0(x)$ be a function on $\manM$ and ${\hat\Omega}_0$
be the BRST charge from Proposition \bref{prop:Omega-special}.
Let also $f(x,Y,\hbar)\in \qA_0$ be the unique solution to the
equation
\begin{equation}
  \qcommut{{\hat\Omega}_0}{f}=0\,,
\end{equation}
with the boundary condition $f{\bigr|}_{Y=0}=f_0\,$ and the additional
conditions $\delta^*f=0\,$ and $f=f(x,Y,\cc,\hbar)$
(see Proposition~\bref{prop:extension-q}).  Then $f$ doesn't depend
on $Y^\alpha$. The unique classical BRST-invariant extension of $f_0$
can be obtained as $f|_{\hbar=0}$ and thus is also
$Y^\alpha$ independent.
\end{prop}
\begin{proof}
The proof goes in the same way as that of Proposition
\bref{prop:Omega-special}.
\end{proof}

Let us write down explicitly a few first terms of the BRST invariant
extension of $f_0(x)$ obtained by
Proposition~\bref{prop:extension-special}:
\begin{multline}
  f=f_0+Y^i\d_if_0+\half Y^iY^j\diff-con^0_i \diff-con^0_j f_0
+\frac{1}{6}Y^i Y^j Y^k\diff-con^0_i \diff-con^0_j \diff-con^0_k f_0
-\frac{1}{24}\cc^i {\bar R}^0_{ij;\,kl}Y^kY^lD^{jm}\d_m f_0+
\cdots\,,
\end{multline}
The expression coincides with that in the Fedosov quantization with
the Poisson bivector $\omega^{ij}$ substituted by the Dirac bivector
$D^{ij}$ and the symplectic connection replaced by the Dirac connection
$\con^0$.

Taking as $f_0$ a constraint function $\theta_\alpha$ one arrives at
the following expression for the unique BRST invariant extension
${\bar\theta}_\alpha$ of $\theta_\alpha$:
\begin{equation}
  {\bar\theta}_{\alpha}=\theta_\alpha+Y^i\d_i\theta_\alpha\,.
\end{equation}
All the higher order terms in the expansion vanish
since $D^{ij}\d_j\theta_\alpha=0$ and
$\d_j\d_i\theta_\alpha-(\con^0)^k_{ij}\d_k\theta_\alpha=0$. 
Consequently one arrives at:
\begin{thm}\label{thm:center}
Let the star-product $\star_\manM$ on $\manM$ is constructed by means of
${\hat\Omega}_0$ from Proposition~\bref{prop:Omega-special}. Then
for any function $f_0$ on $\manM$ one has:
\begin{equation}
f_0\star_\manM \theta_\alpha=
 \theta_\alpha \star_\manM f_0=
f_0\theta_\alpha\,.
\end{equation}
\end{thm}
It follows from the Theorem~\bref{thm:center} that for the star-product
constructed by means of ${\hat\Omega}_0$, the central subalgebra
of the Dirac bracket algebra is also a central subalgebra
of the respective star-commutator algebra.

Thus we obtain the explicit construction of the deformation quantization
of an arbitrary Dirac bracket on a general symplectic manifold,
thereby giving a deformation quantization of respective second-class
constraint system.

\section{Reduction to the constraint surface}\label{sec:reduction}
The purpose of this section is to establish an explicit relation
between the quantization scheme developed in this paper and
the approach based on directly quantizing the respective
second-class constraint surface, considered as a symplectic
manifold without refering to its embedding into the extended phase
space $\manM$.

The goal of the previous sections is to quantize general second-class
system on $\manM$ in terms of the original constraints $\theta_\alpha$
and general coordinates on $\manM$.  In this way we have arrived at
the star-product for the Dirac bracket on $\manM$, which in turn
determines a star-product on the constraint surface $\Sigma$.

On the other hand, one can find quantization of $\Sigma$ as a
symplectic manifold (e.g. using Fedosov approach).  Although
explicit reduction to the constraint surface $\Sigma$ is a huge task
in the realistic physical models it is instructive to trace
the correspondence between the direct quantization of $\Sigma$
as a symplectic manifold and the approach developed above.
Fortunately, it turns out that this approach reproduces not
only the Fedosov star-product on $\Sigma$ but also
all the ingredients (including the extended phase space, the BRST charge,
and the BRST invariant extensions of functions on $\Sigma$)
of the BRST description for the Fedosov
quantization of $\Sigma$.  This, in particular, proves equivalence of
the respective approaches to the constraint system quantization.

\subsection{The extended phase space for $\Sigma$.}
Let us consider the constraint surface $\Sigma$ as
a symplectic manifold, with the symplectic form $\omega^\Sigma
\equiv \omega\bigr|_\Sigma$. Equivalently, the respective Poisson bracket
on $\Sigma$ is the restriction of the Dirac bracket $\pb{\,}{}^D_\manM$ to
$\Sigma$.  We equip $\Sigma$ with the symmetric symplectic connection
$(\con^0)^\Sigma$, which is the restriction of the Dirac connection
$\con^0$ on $\manM$ to $\Sigma$ (see
Section~\bref{subsec:D-connection}).  Then one can apply
to $\Sigma$ the quantization method of Sections~\bref{sec:cl} and
\bref{sec:q}, with $\Sigma$ considered as a phase space of the
unconstraint system (in this case this method reduces
to that of~\cite{[GL]} which in turn provides
a BRST formulation of the Fedosov quantization).
Accordingly, the extended phase space
for $\Sigma$ is given by:
\begin{equation}
\E_\Sigma=\mod(\Pi T\Sigma)\oplus T(\Pi T\Sigma)\,.
\end{equation}
$\E_\Sigma$ is equipped with the symplectic form
\begin{equation}
\begin{split}
  \label{eq:symplectic-Sigma}
\omega^{\E_\Sigma}~~=~~\pi^*(\omega{\bigr|}_\Sigma)~+&~2dp_a\wedge dx^a~+~
2d\cc^a\wedge d\cP_a~+\\
&+~~(\omega{\bigr|}_\Sigma)_{ab}\,dY^a \wedge dY^B
~+~Y^aY^b d\con^0_{ab}
~-~2Y^a \con^0_{ab}\wedge dY^b\,,
\end{split}
\end{equation}
where we have introduced local coordinates
\begin{equation}
  \label{eq:coord-sigma}
  x^a,Y^a,p_a,\cc^a,\cP_a
\end{equation}
on $\E_\Sigma$, with $x^a$ being a local coordinates on $\Sigma$,
considered as function on $\E_\Sigma$, and $Y^a,p_a,\cc^a,\cP_a$
introduced according to Section~\bref{subsec:extended-phase-space}.
Along the line of Sections~\bref{sec:cl} and \bref{sec:q} one can also
identify subalgebras $\qA^\Sigma$ and $\qA^\Sigma_0$,
construct the quantum BRST charge, the quantum BRST observables and find
a covariant star-product on $\Sigma$.  We will see that
all this structures can be obtained by the reduction of the respective
structures from $\E$.

\subsection{Constraints on the extended phase space and reduction to
$\E_\Sigma$ at the classical level.}
A crucial point is that $\E_\Sigma$ can be embedded into the extended phase
space $\E$ of Sections~\bref{sec:cl} and \bref{sec:q}. Indeed,
assume that a connection $\con$ entering the symplectic structure
on $\E$ is the specific connection $\con^0$ given
by~\eqref{eq:con0-explicit} and consider a submanifold of the entire
extended phase space $\E$ determined by the following constraints:
\begin{equation}
\begin{array}{rclrcl}
&&&&&\\[-3pt]
\theta_\alpha&=&0\,, & Y_\alpha&\equiv&Y^i \d_i \theta_\alpha=0\,, \\[4pt]
Y^\alpha&=&0\,,     & p_\alpha&\equiv&p_i\omega^{ij}\d_i\theta_\alpha=0\,, \\[4pt]
\cc^\alpha&=&0\,,   & \cP_\alpha&=&0\,, \\[4pt]
\cc_\alpha&\equiv&\cc^i \d_i \theta_\alpha=0\,, & \qquad
\cP^2_\alpha&\equiv&\cP_i \omega^{ij}\d_j \theta_\beta=0\,.\\[-4pt]
&&&&&
\end{array}
\label{eq:constraints-E}
\end{equation}
This submanifold can be naturally identified with $\E_\Sigma$.
Moreover, the symplectic form~\eqref{eq:symplectic-Sigma} on
$\E_\Sigma$ coincides with the restriction of the
symplectic form $\omega^\E$ defined on $\E$ to $\E_\Sigma$

It is useful to consider $\E_\Sigma$
as a second-class constraint surface in $\E$.
Indeed, a Poisson bracket matrix of constraints
\eqref{eq:constraints-E} reads as:
\vspace{0.5cm}
\begin{equation}
\label{eq:matrix}
  \begin{array}{l|cccccccc}
&\theta_\beta & \cc_\beta&\cc^\beta&Y_\beta&Y^\beta &\cP_\beta &\cP^2_\beta& p_\beta\\
\hline
\theta_\alpha &0&0&0&0&0&0&0&\Delta_{\alpha \beta}\\
\cc_\alpha    &0&0&0&0&0&0&\Delta_{\alpha \beta}              &\bl\\
\cc^\alpha    &0&0&0&0&0&\delta^\alpha_\beta              &\bl&\bl\\
Y_\alpha      &0&0&0&0&-\delta_\alpha^\beta            &\bl&\bl&\bl\\
Y^\alpha      &0&0&0&\delta^\alpha_\beta         &\bl&\bl&\bl&\bl\\
\cP_\alpha    &0&0&\delta^\alpha_\beta        &\bl&\bl&\bl&\bl&\bl\\
\cP^2_\alpha  &0&\Delta_{\alpha\beta}    &\bl&\bl&\bl&\bl&\bl&\bl\\
p_\alpha      &-\Delta_{\alpha\beta}  &\bl&\bl&\bl&\bl&\bl&\bl&\bl
\end{array}
\end{equation}
\vspace{0.5cm}

\noindent
where ``dots'' denote the possibly non-vanishing blocks
whose explicit form is not needed below.
This matrix is obviously invertible.
The matrix, invers to~\eqref{eq:matrix} is given by:
\vspace{0.5cm}
\begin{equation}
\label{eq:DM-invers-E}
  \begin{array}{l|cccccccc}
&\theta_\beta & \cc_\beta&\cc^\beta&Y_\beta&Y^\beta &\cP_\beta &\cP^2_\beta& p_\beta\\
\hline
\theta_\alpha &\bl&\bl&\bl&\bl&\bl&\bl  &\bl &-\Delta^{\alpha\beta}\\
\cc_\alpha    &\bl&\bl&\bl&\bl&\bl&\bl&\Delta^{\alpha\beta} &0\\
\cc^\alpha    &\bl&\bl&\bl&\bl&\bl&\delta_\alpha^\beta&0&0\\
Y_\alpha      &\bl&\bl&\bl&\bl &\delta^\alpha_\beta  &0&0&0\\
Y^\alpha      &\bl&\bl&\bl&-\delta_\alpha^\beta &0&0&0&0\\
\cP_\alpha    &\bl &\bl  &\delta_\alpha^\beta  & 0 &0&0&0&0\\
\cP^2_\alpha  &\bl &\Delta^{\alpha\beta}&0&0&0&0&0&0\\
p_\alpha      &\Delta^{\alpha\beta} &0&0& 0&0&0&0&0
\end{array}
\end{equation}

\begin{prop}\label{prop:Omega-red}
Let $\pb{}{}^D_\E$ be a Dirac bracket associated to
the constraints~\eqref{eq:constraints-E}. Then the BRST charge $\Omega_0$
from Proposition~\bref{prop:Omega-special} satisfies a ``weak'' master
equation:
\begin{equation}
\left(  \pb{\Omega_0}{\Omega_0}^D_\E \right){\bigr|}_{\E_\Sigma}=0\,.
\end{equation}
\end{prop}
\begin{proof}
Let us write down explicitly the following terms:
\begin{equation}
\label{eq:terms}
  \begin{split}
&    \pb{\Omega_0}{\theta_\alpha}_\E=\cc^i\d_i\theta_\alpha=\cc_\alpha\\
&    \pb{\Omega_0}{\cc_\alpha}_\E=0\\
&    \pb{\Omega_0}{\cc^\alpha}_\E=0\\
&    \pb{\Omega_0}{Y_\alpha}_\E=-\cc_\alpha\,.
\end{split}
\end{equation}
The first three equalities are trivial. The last one follows from
Proposition~\bref{prop:Omega-special}.

Further, it follows from the explicit form of the matrix
\eqref{eq:DM-invers-E} entering the Dirac bracket
$\pb{}{}^D_\E$ that each non-vanishing term in $\pb{\Omega_0}{\Omega_0}^D_\E$ is
proportional at least to one quantity from \eqref{eq:terms} and
thereby vanishes on $\E_\Sigma$.
\end{proof}
Similar arguments lead to the following statement:
\begin{prop}\label{prop:observables-red}
Let function $f$ on $\E$ be such that
\begin{equation}
  \pb{\Omega_0}{f}_\E=0\,.
\end{equation}
Let also $f$ satisfies
\begin{equation}
\begin{aligned}
  (\pb{f}{\theta_\alpha}_\E){\bigr|}_{\E_\Sigma}&~=~&0\,,\qquad
  (\pb{f}{Y_\alpha}_\E){\bigr|}_{\E_\Sigma}&~=~&0\,,\\
  (\pb{f}{\cc_\alpha}_\E){\bigr|}_{\E_\Sigma}&~=~&0\,,\qquad
  (\pb{f}{\cc^\alpha}_\E){\bigr|}_{\E_\Sigma}&~=~&0\,.
\end{aligned}
\end{equation}
Then,
\begin{equation}
\label{eq:D-BRST}
  (\pb{\Omega_0}{f}^D_\E){\bigr|}_{\E_\Sigma}=0\,.
\end{equation}
In particular, if $f(x,Y)$ is the unique BRST invariant extension
of the function $f_0(x)$, obtained by
Proposition~\bref{prop:extension-special},
then $f$ satisfies~\eqref{eq:D-BRST}.
\end{prop}

As an obvious consequence of Propositions~\bref{prop:Omega-red}
and \bref{prop:observables-red} we obtain the following relations
\begin{equation}
\pb{\Omega_\Sigma}{\Omega_\Sigma}_{\E_\Sigma}=0\,, \qquad 
\pb{\Omega_\Sigma}{f_\Sigma}_{\E_\Sigma}=0\,,
\end{equation}
where we have introduced separate notations $\Omega_\Sigma$ and
$f_\Sigma$ for the restrictions of the BRST charge $\Omega_0$ and BRST
invariant function $f$ to $\E_\Sigma$. One can also see
that if $f=f(x,Y)$ is a BRST invariant extension of function $f_0(x)$,
then $f_\Sigma=f{\bigr|}_{\E_\Sigma}$ is a BRST invariant
(w.r.t. the BRST charge  $\Omega_\Sigma$) extension of function
$f_0{\bigr|}_{\Sigma}$.

Thus, at the classical level, our scheme reproduces all the
basic structures of the BRST formulation of the Fedosov quantization
for the constraint surface $\Sigma$.  In particular, the extended phase space for
$\Sigma$ is $\E_\Sigma$, the Poisson bracket therein is the
restriction of the Dirac bracket $\pb{}{}^D_\E$ to $\E_\Sigma$, the
BRST charge is $\Omega_\Sigma=\Omega_0{\bigr|}_{\E_\Sigma}$, and the BRST
invariant extension of a function $f_0{\bigr|}_\Sigma\in \func\Sigma$ is the
restriction $f_\Sigma=f{\bigr|}_{\E_\Sigma}$ of the BRST extension $f$
of $f_0\in\func\manM$.

\subsection{Quantum reduction and relation with the Fedosov
quantization of $\Sigma$.} 
Now we are going to show that
the results, analogous to those of the previous subsection,
hold at the quantum level as well.
Let $\qA^\Sigma$ and $\qA^\Sigma_0$ are
the subalgebras of the algebra of functions on $\E_\Sigma$
constructed as in Section~\bref{subsec:quantization} for the unconstraint system
on $\Sigma$.  Let also $\st-sig$ denotes the
quantum multiplication (see Sec.~\bref{subsec:quantization}) in
$\qA^\Sigma$ and $\qA^\Sigma_0$.
\begin{thm}
Let ${\hat\Omega}_\Sigma$ be the restriction of the
quantum BRST charge ${\hat\Omega}_0$ from
Proposition~\bref{prop:Omega-special} to $\E_\Sigma$. Let also $f_\Sigma$ be
a restriction to $\E_\Sigma$ of the quantum BRST invariant
extension $f(x,Y,\hbar)$ obtained by Proposition~\bref{prop:extension-special}
for a function $f_0$ on $\manM$. Then ${\hat\Omega}_\Sigma$ and
$f_\Sigma$ belong to $\qA^\Sigma$ and $\qA^\Sigma_0$ respectively
and satisfy:
\begin{equation}
  \commut{{\hat\Omega}_\Sigma}{{\hat\Omega}_\Sigma}_{\st-sig}=0\,,
\qquad  
  \commut{{\hat\Omega}_\Sigma}{f_\Sigma}_{\st-sig}=0\,.
\end{equation}
\end{thm}
\begin{proof}
Consider the following subset of the constraints
\eqref{eq:constraints-E}:
\begin{equation}
Y_\alpha\equiv Y^i \d_i \theta_\alpha=0\,, \qquad Y^\alpha=0\,,\qquad
\cc^\alpha=0\,,\qquad \cP_\alpha=0\,.
\label{eq:constraints-E-2}
\end{equation}
These constraints are also second-class ones. It is easy to write down
respective Dirac bracket; the non-vanishing basic Dirac bracket
relations are given by:
\begin{equation}
\begin{gathered}
  \label{eq:DB-E}
  \begin{array}{rclrcl}
    \pb{x^i}{p_j}_{\E}&=&\delta^i_j\,, \qquad &
    \pb{Y^i}{Y^j}_{\E}&=&D^{ij}\,,\\[5pt]
    \pb{Y^j}{p_i}_{\E}&=&-(\con^0)^j_{ik}Y^k\,,\qquad &
    \pb{\cc^i}{\cP_j}_{\E}&=&\delta^i_j\,,
  \end{array}\\[4pt]
\pb{p_i}{p_j}_{\E}~~=~~\omega_{ij}+
{\half}{\bar R}^0_{ij;\,kl\,}Y^kY^l\,.
\end{gathered}
\end{equation}
It follows from~\eqref{eq:DB-E} that subalgebras $\qA_0$
and $\qA$ are closed w.r.t. the Dirac bracket \eqref{eq:DB-E}.
Let us introduce the quantum multiplication $\star_D$ in $\qA_0$
build by the restriction of the Dirac bracket~\eqref{eq:DB-E}
to $\qA_0$:
\begin{equation}
\label{eq:Weyl-star-product-D}
(a \star_D b)(x,Y,\cc,\cP,\hbar)= exp ( -\frac{i\hbar}{2}
D^{ij} \dl{Y_1^i} \dl{Y_2^j})
a(x,Y_1,\cc,\cP,\hbar)b(x,Y_2,\cc,\cP,\hbar){\bigr|}_{Y_1=Y_2=Y} \,,
\end{equation}
where $\cP$ stands for dependence on $\cP_\alpha$ only.
\begin{lemma}\label{lemma:2}
\begin{enumerate}
\item
Let $f,g\in \qA_0$ do not depend on $Y^\alpha$ and $\cP_\alpha$. Then,
\begin{equation}
  f\star_D g=f \star g\,,
\end{equation}
where $\star$ in the R.H.S. denotes Weyl
multiplication~\eqref{eq:Weyl-star-product}.
\item For any $a\in \qA_0$ its restriction $a{\bigr|}_{\E_\Sigma}$
belongs to $\qA^\Sigma_0$. Multiplication $\star_D$
determines a multiplication in $\qA^\Sigma_0$
that is identical with $\st-sig$:
\begin{equation}
  (a{\bigr|}_{\E_\Sigma})\st-sig (b{\bigr|}_{\E_\Sigma})=
  ( a \star_D b){\bigr|}_{\E_\Sigma}\,.
\end{equation}
\end{enumerate}
\end{lemma}
\begin{proof}
The only nontrivial is the second statement. It is easy to see that
restriction $f{\bigr|}_{\E_\Sigma}$ of an arbitrary element $f\in \qA_0$
belongs to $\qA^0_\Sigma$. That $\star_D$
restricts to $\qA^\Sigma_0$ follows from the following properties
of $\star_D$:
\begin{equation}
\label{eq:rest-prop}
\begin{split}
  \begin{array}{rccclrcccl}
  f \star_D Y^\alpha&=&Y^\alpha \star_D f&=&f\,Y^\alpha \,,\qquad&
  f \star_D  Y_\alpha& =& Y_\alpha \star_D f&=&f\,Y_\alpha\,,\\[3pt]
  f \star_D C^\alpha&=&\pm C^\alpha  \star_D f& =&f\,C^\alpha \,,\qquad&
  f \star_D  C_\alpha&=&\pm C_\alpha\star_D f & =&f\,C_\alpha\,,\\[3pt]
\end{array}\\
\begin{array}{c}
f\star_D \cP_\alpha= \pm \cP_\alpha\star_D f =
f\,\cP_\alpha\,,\qquad{}\qquad{}\qquad{}\qquad{}\qquad{}
\end{array}\\
\end{split}
\end{equation}
for any $f\in \qA_0$. Indeed, $\qA_0^\Sigma$
(as an algebra w.r.t. ordinary commutative product) can be identified with the quotient of
$\qA_0$ modulo ideal generated (by ordinary commutative product)
by elements
$Y^\alpha,Y_\alpha,\cc^\alpha,\cc_\alpha,\cP_\alpha\in \qA_0$.  Then,
\eqref{eq:rest-prop} implies that this
ideal is also an ideal in $\qA_0$ w.r.t. the $\star_D$-product.
Thus $\star_D$ determines a star-product in $\qA^\Sigma_0$.  One can
easily check that in $\qA_0^\Sigma$ the product coincides with
$\st-sig$.
\end{proof}
Let us now rewrite the master equation~\eqref{eq:master-q}
for ${\hat\Omega}_0$ in the
``Fedosov'' form:
\begin{equation}
-\delta r + \Dcon-d r+\frac{i}{2\hbar}\commut{r}{r}_\star={\bar R}^0\,,
\end{equation}
where
\begin{equation}
r=\sum_{s=3}^\infty {\hat\Omega}^s_0\,, \qquad 
\Dcon-d =\cc^i\d_i-\cc^i (\con^0)^A_{iB}Y^B\dl{Y^A}\,, \qquad
\delta=\cc^A\dl{Y^A}\,,
\end{equation}
and
\begin{equation}
  {\bar R}^0=
\frac{1}{4} \cc^i \cc^j  {\bar R}^0_{ij;\, A B} Y^A Y^B=
\frac{1}{4} \cc^i \cc^j  {\bar R}^0_{ij;\, kl} Y^k Y^l \,.
\end{equation}
Since $r$ doesn't depend on $Y^\alpha$ the master equation can be
equivalently rewritten using multiplication $\star_D$
and the Dirac connection on $\manM$:
\begin{equation}
-\delta r+(\cc^i\d_i-
\cc^i (\con^0)^k_{ij}Y^j\dl{Y^j})r+
\frac{i}{2\hbar}\commut{r}{r}_{\star_D}={\bar R}^0\,.
\end{equation}
Taking into account the second item of Lemma~\bref{lemma:2}
one arrives at
\begin{equation}
\label{eq:F-master-sigma}
-\delta_{\Sigma}(r{\bigr|}_{\E_\Sigma})
+{\Dcon-d}_{\Sigma}(r{\bigr|}_{\E_\Sigma})
+\frac{i}{2\hbar}\commut{r{\bigr|}_{\E_\Sigma}}{r{\bigr|}_{\E_\Sigma}}_{\st-sig}
={\bar R}^0{\bigr|}_{\E_\Sigma}\,.
\end{equation}
where $\Dcon-d_{\Sigma}$ and $\delta_{\Sigma}$ are restrictions
of $\Dcon-d$ and $\delta$ defined on $\E$ to ${\E_\Sigma}$. In the
coordinates $x^a,p_a,Y^a,\cc^a,\cP_a$ on $\E_{\Sigma}$ one has
\begin{equation}
{\Dcon-d}_{\Sigma}=\cc^a\d_a-\cc^a (\con^0)^c_{ab}Y^b\dl{Y^c}\,,
\qquad \delta_{\Sigma}=\cc^a \dl{Y^a}\,,
\end{equation}
where $(\con^0)^c_{ab}$ are coefficients of a connection $(\con^0)^\Sigma$ on
$\Sigma$  w.r.t. the coordinates $x^a$ (recall that Dirac connection
$(\con^0)$ on $\manM$ restricts to $\Sigma$).

Finally, one can observe that
for
\begin{equation}
{\hat\Omega}_\Sigma\equiv{\hat\Omega_0}{\bigr|}_{\E_\Sigma}=
-\cc^ap_a-\cc^a \omega^\Sigma_{ab}y^b+r{\bigr|}_{\E_\Sigma}
  \end{equation}
Eq.~\eqref{eq:F-master-sigma} implies:
\begin{equation}
  \commut{{\hat\Omega}_\Sigma}{{\hat\Omega}_\Sigma}_{\st-sig}=0\,.
\end{equation}
Similar arguments show the rest of the statement.
\end{proof}
It follows from the theorem that the star product on $\Sigma$ obtained
in Section~\bref{subsec:BRST-q} as a quantum multiplication
of the nonequivalent quantum observables can be identified with
the Fedosov star product on $\Sigma$, provided
the symplectic connection entering the Poisson bracket on $\E$
is the specific symplectic connection $\con^0$; the Fedosov star
product on $\Sigma$ corresponds then to the connection
$(\con^0)^\Sigma$ on $\Sigma$
obtained by the restriction of the Dirac connection $\con^0$
defined on $\manM$ to $\Sigma$.

\section{An alternative formulation}\label{sec:alternative}
An interesting understanding of the quantization scheme proposed in
this paper is provided by considering of the symplectic form
$D$ in the vector bundle $\W(\manM)$ as the symplectic form on
the appropriate symplectic manifold.  Namely, let us consider the vector
bundle $\V(\manM)=\manM \times V$ associated with the constraints
$\theta_\alpha$ (see the beginning of
Section~\bref{subsec:connection}).  Let $\eta^\alpha$ be coordinates on
$V$.  Considered as a manifold $\V(\manM)$ is equipped with the
following 2-form:
\begin{equation}
  \label{eq:D-form}
  \aD=\pi^*\omega-d \theta_\alpha \wedge d \eta^\alpha\,.
\end{equation}
It is useful to write respective matrix:
\begin{equation}
\aD=
\left(\begin{array}{cc}
\aD_{ij} & \aD_{i\beta}\\[7pt]
\aD_{\alpha j}& \aD_{\alpha\beta}
 \end{array}\right)=
\left(\begin{array}{cc}
\omega_{ij} & \d_i \theta_\beta\\[7pt]
- \d_j\theta_\alpha & 0
 \end{array}\right)
\end{equation}
The 2-form $\aD$ is closed and nondegenerate,
provided the respective Dirac matrix
$\Delta_{\alpha\beta}=\pb{\theta_\alpha}{\theta_\beta}_\manM$
is invertible.  Introducing a unified notation $x^A$ for the
coordinates $x^i$ and $\eta^\alpha$ it is easy to see that
coefficients $\aD_{AB}$ of the symplectic form $\aD$ coincides with the
coefficients of the symplectic form $D$ on the fibres of $\W(\manM)$ from
Section~\bref{sec:cl} (it is assumed that the coefficients corresponds
to the same coordinate system on $\manM$ and the same basis of
constraints).  Speaking geometrically, the fibres of
the symplectic vector bundle $\W(\manM)$ is identified with the
fibres of the tangent bundle over $\V(\manM)$.

The Poisson bracket corresponding to the symplectic form
\eqref{eq:D-form} reads as
\begin{equation}
  \label{eq:D-PB}
\begin{split}
\pb{x^i}{x^j}_{\V(\manM)}&=\omega^{ij}-\omega^{il}(\d_l \theta_\alpha)
\Delta^{\alpha \beta} (\d_k \theta_\beta) \omega^{kj}\,, \\
\pb{x^i}{\eta^\alpha}_{\V(\manM)}&=-\omega^{il}(\d_l \theta_\beta)
\Delta^{\beta \alpha}\,,\\
\pb{\eta^\alpha}{\eta^\beta}_{\V(\manM)}&=\Delta^{\alpha\beta}\,.
\end{split}  
\end{equation}
Note that this nondegenerate Poisson bracket coincides with the Dirac
bracket on $\manM$ when evaluated on $\eta$-independent functions.
Also, the respective Poisson bracket matrix coincides with the
matrix $D^{AB}$ \eqref{eq:DAB}.

\subsection{First class constraint system on $\V(\manM)$}
Let us consider functions $\theta_\alpha$ as the constraints on
$\V(\manM)$.  It is easy to see that these constraints are the first
class ones.  Moreover, the first-class system determined by
$\theta_\alpha$ is Abelian. Indeed,
\begin{equation}
  \pb{\theta_\alpha}{\theta_\beta}_{\V(\manM)}=
  \pb{\theta_\alpha}{\theta_\beta}^D_{\manM}=0\,,
\end{equation}
where  $\pb{}{}^D_{\manM}$ is a Dirac bracket on $\manM$
associated with the second-class constraints $\theta_\alpha$.  As a
matter of simple analysis this first-class system is
equivalent to the original second-class system on $\manM$.

This representation of the second-class system on $\manM$
allows one to develop an alternative quantization procedure
based on quantizing the first-class system on $\V(\manM)$. However,
to quantize this first-class system one should first find quantization
of $\V(\manM)$ considered as a symplectic manifold.

\subsection{BRST quantization of $\V(\manM)$}\label{subsec:V-quant}
The quantization of $\V(\manM)$ is rather standard and can be obtained
within the quantization scheme of Section~\bref{sec:q} applied to the
unconstraint system on $\V(\manM)$ (in this case the approach
reduces to that proposed in~\cite{[GL]} and results in the Fedosov
star-product on $\V(\manM)$).

According to the scheme one should fix a symmetric symplectic
connection in the tangent bundle $T\V(\manM)$.  In fact we already
have this connection. Indeed, because $T\V(\manM)$ can be identified
with the vector bundle $\W(\manM)$ from Section~\bref{sec:cl}, pulled back by
the projection $\V(\manM) \to \manM$, the connection in $T\V(\manM)$ can be
obtained from that in $\W(\manM)$. Namely, let us consider a
connection on $\V(\manM)$ determined by
\begin{equation}
\label{eq:conn-explicit}
\conn^A_{iB}={\D-con}^A_{iB}\,, \qquad \conn^A_{\alpha B}=0\,,
\end{equation}
where coefficients ${\D-con}^A_{iB}$ are given
by~\eqref{eq:con0-explicit}.
One can easily check that $\conn$ is indeed a
symmetric symplectic connection on $\V(\manM)$.  Note that
coefficients of $\conn$ in the coordinate system $x^i,\eta^\alpha$ on
$\V(\manM)$ do not depend on $\eta^\alpha$.

Let $\bar\E$ be the extended phase space of the unconstraint system
on $\V(\manM)$, which is  constructed according to
Section~\bref{subsec:extended-phase-space}. Let also 
$p_A,\,Y^A,\,\cc^A$ and $\cP^A$ denote momenta, conversion variables, ghost
variables, and ghost momenta respectively.
In this setting Poisson bracket~\eqref{eq:PB-E} reads as
\begin{equation}
\begin{gathered}
  \label{eq:PB-E-2}
 \begin{array}{rclrcl}
    \pb{x^A}{p_B}_{\bar\E}&=&\delta^A_B\,, \qquad&
    \pb{Y^A}{Y^B}_{\bar\E}&=&\aD^{AB}\,,\\[5pt]
    \pb{Y^A}{p_B}_{\bar\E}&=&-\conn^A_{BC}Y^B\,,\qquad&
    \pb{\cc^A}{\cP_B}_{\bar\E}&=&\delta^A_B\,, 
  \end{array}\\[5pt]
  \pb{p_A}{p_B}_{\bar\E}=
\aD_{AB}+{\half}{\tilde {\bar R}}_{AB;\,CD\,}Y^C Y^D\,,
\end{gathered}
\end{equation}
where $\tilde {\bar R}$ denotes a curvature of the connection $\conn$
on $\V(\manM)$.

Specifying the constructions of Section~\bref{sec:q}
to the extended phase space $\bar \E$ one obtains
a unique quantum BRST charge $\hat{\bar\Omega}\in\qA^{\bar\E}$
satisfying
\begin{equation}
  \qcommut{\hat{\bar\Omega}}{\hat{\bar\Omega}}=0\,, \qquad
\p{\hat{\bar\Omega}}=1\,,\quad \gh{\hat{\bar\Omega}}=1\,,
\end{equation}
the boundary condition
\begin{equation}
\hat{\bar\Omega}^0=0\,,\qquad
\hat{\bar\Omega}^1=-\cc^A D_{AB} Y^B\,,\qquad
\hat{\bar\Omega}^2=-\cc^A p_A\,,
\end{equation}
and the additional conditions conditions
${\hat{\bar\Omega}}^r \in \qA^{\bar\E}_0$
and $\delta^* \hat{\bar\Omega}^r=0$ for $r\geq 3$.

A unique quantum BRST extension of a function $f_0(x^A)$ is the
solution to the equation
\begin{equation}
  \qcommut{{\hat{\bar\Omega}}}{f}=0\,, \qquad f|_{Y=0}=f_0\,,
 \quad f\in\qA^{\bar\E}_0\,,\quad \gh{f}=\gh{f_0}\,,
\end{equation}
subjected to the additional condition $\delta^* f=0$.

In this way one can find the star product $\star_{\V(\manM)}$ on
$\V(\manM)$, giving a deformation quantization of the Poisson bracket
\eqref{eq:D-PB} on $\V(\manM)$.  An important point is that
${\hat{\bar\Omega}}^r$ doesn't depend on the variables $\eta^\alpha$ and
$p_\alpha$ for $r\geq 3$.  The same holds for the unique BRST
invariant extension of a function $f_0(x^i)$.  Since the quantum
multiplication in $\qA^{\bar\E}$ obviously preserve the space of
$\eta^\alpha$ and $p_\alpha$ independent elements, the star product
$\star_{\V(\manM)}$ preserves the space of $\eta$-independent
functions, giving thus a deformation quantization of the Dirac bracket on
$\manM$. Given $\eta$-independent functions $f_0(x)$ and $g_0(x)$ one
has
\begin{equation}
\label{eq:star-a}
  f_0\star_\manM g_0=f \star g |_{Y=0}=f_0
  g_0-\frac{i\hbar}{2}\pb{f_0}{g_0}^D_{\manM}+\cdots\,,
\end{equation}
where $f$ and $g$ are the unique quantum BRST extensions of $f_0$ and
$g_0$, $\star$ is the Weyl product in $\qA^{\bar\E}_0$, and $\cdots$
denote higher order terms in $\hbar$.

\subsection{The total BRST charge}
In spite of the fact that the BRST charge $\hat{\bar\Omega}$
constructed above allows one to find a star product for the Dirac
bracket on $\manM$, the BFV-BRST theory determined by $\hat{\bar\Omega}$ is
not equivalent to the original second-class system on $\manM$.
The matter is that the original first-class constraints $\theta_\alpha$ on
$\V(\manM)$ have not been taken into account.

A way to incorporate the original first-class constraints is well
known~\cite{[BFF],[BT]}. To this end one should find
BRST invariant extensions of the original first-class
constraints and then incorporate them into the appropriately
extended BRST charge with their own ghost variables.

Specifying the construction to the case at hand let ${\bar\cc}^\alpha$
and ${\bar\cP}_\alpha$ be the ghosts and their conjugate momenta
associated to the first-class constraints $\theta_\alpha$.
A total BRST charge ${\hat{\bar\Omega}}_{total}$
is then given by
\begin{equation}
  {\hat{\bar\Omega}}_{total}={\hat{\bar\Omega}}+{\bar\cc}^\alpha{\bar\theta}_\alpha=
{\hat{\bar\Omega}}+{\bar\cc}^\alpha{\theta}_\alpha
+{\bar\cc}^\alpha Y^i\d_i\theta_\alpha\,.
\end{equation}
Because $\bar\theta_\alpha$ is the BRST invariant extension of
$\theta_\alpha$, ${\hat{\bar\Omega}}_{total}$ is obviously nilpotent.
Finally, one can check that ${\hat{\bar\Omega}}_{total}$ determines
the correct spectrum of observables.

\subsection{Equivalence to the standard approach}
To complete the description of the alternative formulation we show
that the star product~\eqref{eq:star-a} on $\manM$ coincides with that
obtained in Section~\bref{subsec:special}.

Note, that the extended phase space $\E$
constructed in Section~\bref{subsec:extended-phase-space}
can be identified with the submanifold in $\bar\E$ determined by
$\eta^\alpha=p_\beta=0$.  Since ${\hat{\bar\Omega}}^r$
do not depend on the variables $\eta^\alpha$ and $p_\beta$
for $r \geq 3$, they can can be considered as functions on $\E$.
\begin{prop}
Let $\hat{\bar\Omega}$ be the unique quantum BRST charge
of the unconstraint system on $\V(\manM)$
obtained in Section~\bref{subsec:V-quant}
and ${\hat\Omega}_0$ be the unique quantum BRST charge of the second
class system on $\manM$ obtained in
Proposition~\bref{prop:Omega-special}. Then
\begin{equation}
\label{eq:add}
  {\hat{\bar\Omega}}^r={\hat\Omega}_0^r\,, \qquad r \geq 3\,.
\end{equation}
where  ${{\hat{\bar\Omega}}}^r$ and ${{\hat\Omega}_0}^r$
are the respective terms in the expansions of ${{\hat{\bar\Omega}}}$ and
${{\hat\Omega}}$ w.r.t. degree:
$\deg{{\hat\Omega}^r_0}=\deg{{\hat{\bar\Omega}}^r}=r$.
\end{prop}
\begin{proof}
The statement of the theorem can be explicitly checked for $r=3$.
Further, assuming that~\eqref{eq:add} holds for all $r \leq p$
one can see that the respective quantities ${\hat B}^p$
and ${\hat{\bar B}}^p$ (see the proof of
Theorem~\bref{thm:existence-q}) do coincide.
Since the operators $\delta$ and $\delta^{*}$ are precisely the same
in both cases one observes that
${\hat{\bar\Omega}}^{p+1}={\hat\Omega}^{p+1}$. The
statement then follows by induction.
\end{proof}
It follows from the theorem that for
an arbitrary function $f$ depending on $x^i,Y^i$ and $\hbar$ only
one has $\qcommut{\hat{\bar\Omega}}{f}=\qcommut{{\hat{\Omega}}_0}{f}$.
This implies that the unique BRST invariant extensions
of functions from $\manM$, determined by
${{\hat{\bar\Omega}}}$ and
${{\hat\Omega}_0}$ do coincide. This, in turn, implies that
the star product on $\manM$ given by~\eqref{eq:star-a} coincides
with that obtained in Section~\bref{subsec:special} using
the BRST charge $\hat\Omega_0$.

As a final remark we note that the equivalence statement can also be
generalized to the case where connection $\con$ entering the
symplectic structure on $\E$ is an arbitrary symplectic connection in
$\W(\manM)$. In this setting, however, one should equip $T\V(\manM)$
with the symplectic connection appropriately build by~$\con$.

\section{Conclusion}
We summarize the results of this paper.  For a second-class constraint
system on an arbitrary symplectic manifold $\manM$, we have
constructed an effective first-class constraint (gauge) system
equivalent to the original second-class one. The construction is based
on representing the symplectic manifold as a second-class surface in
the cotangent bundle $\mod\manM$ equipped with a modified symplectic
structure.  The second-class system on $\mod\manM$ determined by the
constraints responsible for the embedding of $\manM$ and the original
second-class constraints are converted into an effective first-class
system by applying a globally defined version of the standard
conversion procedure.  Namely, the conversion variables are introduced
as coordinates on the fibres of the vector bundle
$\W(\manM)=T\manM\oplus \V(\manM)$ associated with the complete set
of constraints, with the symplectic form given by the respective Dirac
matrix.  The phase space of the effective system is equipped with the
specific symplectic structure build with the help of a symplectic
connection in $\W(\manM)$.  We present an explicit form of the
particular symplectic connection in $\W(\manM)$, which is in some
sense a minimal one.  Remarkably, this connection reduces to the Dirac
connection on~$\manM$, i.e., to a symmetric connection compatible with
the Dirac bracket on~$\manM$.

The effective gauge system thus constructed is quantized by the
BFV--BRST procedure. The respective algebra of quantum BRST observables is
explicitly constructed and the star product for the Dirac bracket
on~$\manM$ is obtained as the quantum multiplication of the BRST observables.

In the case where the effective gauge system is constructed by the
particular (Dirac) connection, the original second-class constraints
are shown to be in the center of the respective star-commutator
algebra.  When restricted to the constraint surface, this star product
is also shown to coincide with the Fedosov product constructed by
restricting the Dirac connection to the surface.

The proposed quantization method is explicitly phase space covariant
(i.e., covariant with respect to the change of local coordinates on
the phase space) and doesn't require solving the constraint equations
(which is usually impossible in physically relevant theories).
An interesting problem is to construct the generalization of this
method that is also covariant under changing the basis of
constraints.

\subsection*{Acknowledgments}
We wish to thank V.A.~Dolgushev, A.V.~Karabegov, A.M.~Semikhatov,
A.A.~Sha\-ra\-pov, I.Yu.~Tipunin and I.V.~Tyutin for
the discussions of different problems related to this work.
This work is supported by INTAS-00-262. The waork of I.A.B
is supported by RFBR grant 99-01-00980. The work of M.A.G is partially
supported by the RFBR grant 98-01-01155, Russian Federation President
Grant~99-15-96037 and Landau Scholarship Foundation, Forschungszentrum
J\"ulich.  The work of S.L.L is supported by the RFBR grant
00-02-17956.

\end{document}